%% file: Main.tex
\documentclass[aps,
prb,
preprint,
superscriptaddress,
groupedaddress,
amsmath,amssymb,
floatfix,
12pt, 
]{revtex4}
\usepackage{graphicx}
\usepackage{wrapfig}
\usepackage{dcolumn}
\usepackage{bm}
\usepackage{seqsplit} 
\usepackage{ulem}     
\usepackage{float}
\usepackage[monochrome]{xcolor} 
\usepackage{mathdots} 
\usepackage{subfiles}

\begin{document}


\title{Role of Acoustic Phonon Transport in Near- to Asperity-Contact Heat Transfer}

\author{Amun Jarzembski}
\thanks{A.J. and T.T. contributed equally to this work.}
\affiliation{
Department of Mechanical Engineering, University of Utah, Salt Lake City, Utah 84112, United States.}
\affiliation{
Sandia National Laboratories, Albuquerque, New Mexico 87185, United States.
}

\author{Takuro Tokunaga}
\thanks{A.J. and T.T. contributed equally to this work.}
\affiliation{
Department of Mechanical Engineering, University of Utah, Salt Lake City, Utah 84112, United States.}

\author{Jacob Crossley}
\affiliation{
Department of Mechanical Engineering, University of Utah, Salt Lake City, Utah 84112, United States.}

\author{Jeonghoon Yun}
\affiliation{
Department of Mechanical Engineering, Korea Advanced Institute of Science and Technology, Daejeon 34141, South Korea.}

\author{Cedric Shaskey}
\affiliation{
Department of Mechanical Engineering, University of Utah, Salt Lake City, Utah 84112, United States.}

\author{Ryan A. Murdick}
\affiliation{
Renaissance Scientific, Boulder, CO 80301, United States.}

\author{Inkyu Park}
\affiliation{
Department of Mechanical Engineering, Korea Advanced Institute of Science and Technology, Daejeon 34141, South Korea.}
 
\author{Mathieu Francoeur}
\email{mfrancoeur@mech.utah.edu}
\affiliation{
Department of Mechanical Engineering, University of Utah, Salt Lake City, Utah 84112, United States.}

\author{Keunhan Park}
\email{kpark@mech.utah.edu}
\affiliation{
Department of Mechanical Engineering, University of Utah, Salt Lake City, Utah 84112, United States.}


\begin{abstract}
Acoustic phonon transport is revealed as a potential radiation-to-conduction transition mechanism for single-digit nanometer vacuum gaps. To show this, we measure heat transfer from a feedback-controlled platinum nanoheater to a laterally oscillating silicon tip as the tip-nanoheater vacuum gap distance is precisely controlled from a single-digit nanometer down to bulk contact in a high-vacuum shear force microscope. The measured thermal conductance shows a gap dependence of $d^{-5.7\pm1.1}$ in the near-contact regime, which is in good agreement with acoustic phonon transport modeling based on the atomistic Green's function framework. The obtained experimental and theoretical results suggest that acoustic phonon transport across a nanoscale vacuum gap can be the dominant heat transfer mechanism in the near- and asperity-contact regimes and can potentially be controlled by an external force stimuli. 
\end{abstract}

\maketitle



\subfile{Manuscript.tex}

\newpage
\clearpage

\renewcommand{\thefigure}{S\arabic{figure}}
\setcounter{equation}{0}

\subfile{PRB_SI.tex}

\end{document}

%% file: Manuscript.tex
\section{Introduction}
Heat transfer between bodies separated by nanoscale vacuum gap distances has been extensively studied for potential applications in thermal management \cite{Fiorino2018a,Zhu2019}, energy conversion \cite{Fiorino2018,Inoue2019,Bhatt2020,Lucchesi2021} and data storage \cite{Albisetti2016}. For vacuum gap distances down to 10 nm, state-of-the-art experiments demonstrated that heat transport is mediated by near-field radiative heat transfer (NFRHT), which can greatly exceed Planck's blackbody limit \cite{Rousseau2009a,Shen2009,Song2015,St-gelais2016,Bernardi2016,Song2016,Watjen2016,Ghashami2018,Lim2018,Kim2015a,DeSutter2019,Tang2020}. By comparing measurements with theory based on fluctuational electrodynamics \cite{Polder1971,Rytov1989}, the tunneling of evanescent electromagnetic waves has been unambiguously identified as the enhancement mechanism. In contrast, phonons become the dominant heat carrier when two objects are brought into contact \cite{Gotsmann2013}. This suggests that there should be a transition between electromagnetic wave-mediated NFRHT and phonon-mediated heat conduction in the near-contact regime.

Various theoretical studies have explored acoustic phonon transport across single-digit nanometer vacuum gaps as a plausible radiation-to-conduction transition mechanism \cite{Sellan2012, Xiong2014, Chiloyan2015b, Alkurdi2020,Prunnila2010, Ezzahri2014, Pendry2016, Wang2017, Sasihithlu2017, Zhang2018, Venkataram2020,Volokitin2020, Chen2021,Tokunaga2021,Tokunaga2022}. These efforts have highlighted the roles of interatomic \cite{Sellan2012, Xiong2014, Chiloyan2015b, Alkurdi2020,Prunnila2010, Ezzahri2014, Pendry2016, Wang2017, Sasihithlu2017, Zhang2018, Venkataram2020, Chen2021,Tokunaga2021,Tokunaga2022} and electrically-driven force interactions \cite{Pendry2016,Volokitin2020,Tokunaga2021} in mediating the so-called acoustic phonon tunneling phenomenon. Only one study experimentally explored acoustic phonon tunneling in the near-contact regime by implementing a scanning tunneling microscope with inelastic electron tunneling spectroscopy for thermal measurement \cite{Altfeder2010}. However, this work employed a simple photon emission model to support the presence of acoustic phonon tunneling and lacks rigorous comparison with measurements. Other studies have probed heat transfer for sub-10-nm vacuum gap distances \cite{Kim2015a,Kittel2005a,Kloppstech2017,Cui2017a}, but have not linked heat transfer measurements across both the gap and contact regimes with a unified theoretical model to elucidate the existence of gap-mediated acoustic phonon transport.

This article presents experimental and theoretical results demonstrating that acoustic phonon transport can dominate heat transfer in the near- to asperity contact regimes. To this end, we measure thermal transport from a feedback-controlled platinum (Pt) nanoheater to a flattened silicon (Si) tip in a high-vacuum shear force microscopy (HV-SFM) platform, which can precisely control the tip-nanoheater vacuum gap from single-digit nanometers to bulk contact. By selecting dissimilar materials (Pt and Si), NFRHT is effectively suppressed to make acoustic phonon transport the dominant heat transfer mechanism over the gap range considered\cite{Song2015}. 
Each of the experimental results are quantitatively compared with calculations based on the atomistic Green's function (AGF) method for acoustic phonon transport and fluctuational electrodynamics for NFRHT. 
The theoretical predictions indicate that acoustic phonon transport is driven by the Coulomb force interaction for near-contact vacuum gaps, which gives way to strong interatomic forces at the onset of contact. By analyzing the simultaneously measured tip-nanoheater thermal conductance and lateral force interaction, direct proportionality consistent with the AGF method is revealed and further emphasizes the role of acoustic phonon transport in the experiment. The results indicate acoustic phonon transport as a potential radiation-to-conduction transition mechanism, which can be used to develop active nanoscale thermal management systems.

\section{HV-SFM/Nanoheater Experiments}
\subsection{Experimental setup}

\begin{figure} [t!]
  \begin{center}
    \includegraphics[width=0.65\linewidth]{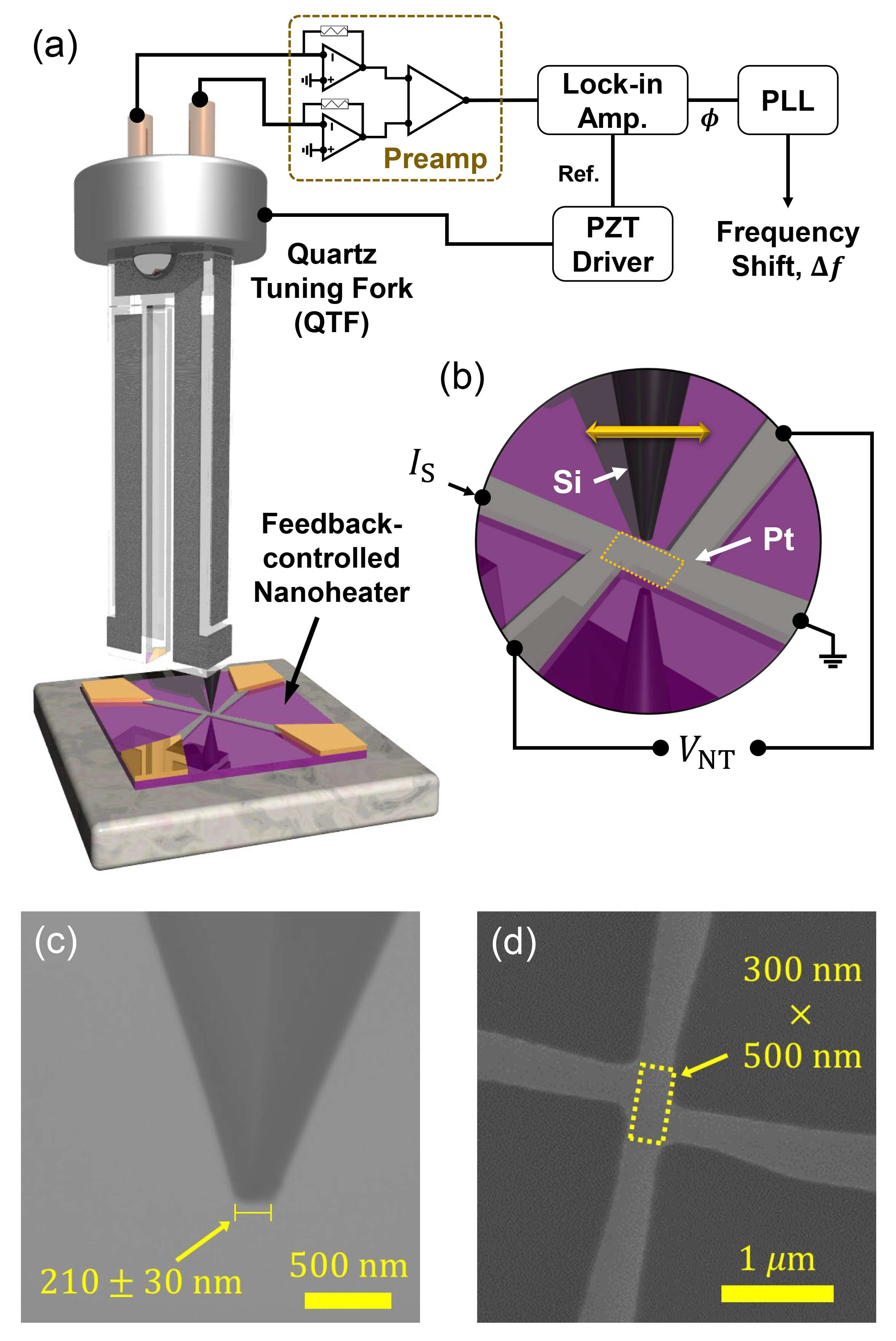}
  \end{center}
  \caption{(a) Experimental schematic based on a custom-built HV-SFM and feedback-controlled Pt nanoheaters. (b) Electrical schematic of the nanoheater four-point probe detection scheme with a laterally-oscillating Si tip in close proximity. (c) SEM image of the Si tip, which exhibits a flat top geometry whose width is 210 $\pm$ 30 nm. (d) SEM image of the nanoheater device showing its maximum sensing area size of 300 nm $\times$ 500 nm.}
  \label{Fig:Schematic}
\end{figure}

The HV-SFM, shown in Fig. \ref{Fig:Schematic}(a), is a custom-built, high vacuum system (5$\times$10$^{-6}$ Torr in routine operations) that adopts a vertically aligned quartz tuning fork (QTF) probe having an Si tip at the free end of one prong for sensitive tip-sample lateral force measurement. \textcolor{red}{The HV-SFM is equipped with a piezoelectric sample stage having an \textit{x,y,z}-scanning range of 30 $\mu$m $\times$ 30 $\mu$m $\times$ 10 $\mu$m, respectively, and 16-pin electrical feedthroughs for electric connections during vacuum experiments. The vertical displacement of the sample stage is carefully calibrated to reveal a $z-$piezo sensitivity of 28.0$\pm$2.8 nm/V with the position uncertainty of 1.8 \AA \hspace{1mm} \cite{SI_Jarzembski}, which is precise enough to control the vertical position of the tip with a sub-nanometer resolution.}

When mechanically driven at the in-plane, anti-symmetric resonant frequency ($f_{0}$ = 32.768 kHz), the QTF probe exhibits a quality factor of $\sim$4000 in a high-vacuum environment. This high quality factor allows single-angstrom resolution for tip-sample gap control through monitoring the QTF's resonance frequency shift (i.e., $\Delta f= f - f_{0}$) \cite{Karrai2000}. According to the first-order QTF oscillator model, its resonance frequency shift enables quantification of the tip-sample conservative lateral force graidient ($\partial F_{x}/\partial z$) using $\partial F_{x}/\partial z \approx 2k_{\mathrm{eff}} \Delta f/f_0$ \cite{Castellanos-Gomez2011}. 
Here, the effective spring constant, $k_{\mathrm{eff}}$, can be approximated as $k_{\mathrm{eff}} \approx Ewt^{3}/(4l^{3})$, where $E = 7.87\times10^{10}\ \mathrm{N/m^2}$ is the Young's modulus of quartz \cite{Grober2000}. Using the dimensions of the QTF prong ($l = 3.52$ mm; $w = 0.25$ mm; $t = 0.58$ mm), $k_{\mathrm{eff}}$ is estimated to be 22.0 kN/m. In addition, the vertically mounted QTF probe enables tip position stability above the sample surface with sub-nanometer gap control, which is not possible in the conventional cantilever-based method due to snap-in contact \cite{Kim2015a}. HV-SFM is also advantageous over the electron tunneling methodologies \cite{Altfeder2010,Kittel2005a,Kloppstech2017,Cui2017a} as the QTF-mounted tip is not necessarily limited to electrically conductive materials. For the present study, an Si tip modified to a flat-top, 210 $\pm$ 30 nm in width as shown in Fig. \ref{Fig:Schematic}(c), is used to secure a sufficiently large heat transfer area as well as to implement a plane-plane configuration for theoretical modeling.

\textcolor{red}{In order to minimize any effect of lateral tip motion onto the tip-nanoheater thermal transport measurement, the QTF oscillation amplitude is maintained within the sub-nanometer range. The lateral oscillation amplitude of a QTF probe was measured by optical fiber interferometry (OFI). As illustrated in Fig. S2(a) of the Supplemental Material \cite{SI_Jarzembski}, the optical fiber aperture is aligned to the side wall of the tip-attached QTF prong to measure its lateral oscillation amplitude. The lateral oscillation amplitude of the tip ($\Delta x_{\mathrm{tip}}$) is then estimated from the OFI measurement by using the QTF geometry. Both the electrical outputs from the QTF and OFI are simultaneously demodulated at the QTF resonant frequency to correlate $\Delta x_{\mathrm{tip}}$ with the QTF electrical signal ($\Delta x_{\mathrm{e}}$). Figure S2(b) of the Supplemental Material \cite{SI_Jarzembski} shows a linear correlation between $\Delta x_{\mathrm{tip}}$ and $\Delta x_{\mathrm{e}}$, from which the QTF amplitude signal sensitivity is determined to be $7.32 \pm 0.05$ nm/V. The QTF electrical signal is set to $\sim$70 mV-rms at its resonance frequency during experiments, which corresponds to a lateral tip motion of $\sim$0.5 nm-rms. This lateral tip motion is on the order of the average lattice constant of the Si-Pt system \cite{Feibelman2001,Esfarjani2011} and is approximately three orders of magnitude smaller than the effective surface area subjected to thermal transport. Such a small lateral tip oscillation does not affect the thermal transport measurement.}

The nanoheaters are batch fabricated using e-beam lithography for the nanopatterned Pt strip and photolithography for the micro-patterned gold electrodes \cite{Hamian2016b}. The Pt nanoheater has a sensing region of approximately 300 nm $\times$ 500 nm between the two inner electrodes, as marked by a yellow dashed box in Figs. \ref{Fig:Schematic}(b) and (d), allowing a four-point probe electrical resistance measurement. When an electrical current ($I_{\mathrm{S}}$) is supplied to the nanoheater for Joule-heating, the voltage drop ($V_{\mathrm{NT}}$) across the inner electrodes is measured to monitor the electrical resistance of the nanoheater's sensing region ($R_{\mathrm{NT}}$). \textcolor{red}{For calibration, a nanoheater chip is placed on a heater stage equipped with a temperature controller (Cryo-Con, Model 22C) in the HV-SFM vacuum chamber. The sensing current is set to $I_{\mathrm{S}} =$ 100 $\mu$A to minimize self-heating, while the entire nanoheater chip is bulk-heated in a high vacuum condition \cite{Jarzembski2018}. Figure S3 in the Supplemental Material shows the calibration results of two nanoheaters (nanoheater \#1 for the near-contact measurements and nanoheater \#2 for the bulk-contact measurements)\cite{SI_Jarzembski}, demonstrating that $R_{\mathrm{NT}}$ is linearly proportional to the substrate temperature $T_{\mathrm{S}}$ (or the sensing area temperature $T_{\mathrm{NT}}$). The resultant temperature coefficient of resistance (TCR) is 1.2$\times$10$^{-3}$ K$^{-1}$ for nanoheater \#1 and 9.6 $\times$ 10$^{-4}$ K$^{-1}$ for nanoheater \#2, respectively. Moreover, the sensing region of the nanoheater can be Joule-heated up to $\sim$500 K by increasing power dissipation (i.e., $P_{\mathrm{NT}} = I_{\mathrm{S}} \times V_{\mathrm{NT}}$). From the obtained linear correlation between $T_{\mathrm{NT}}$ and $P_{\mathrm{NT}}$, the effective thermal resistance of the sensing region ($R_{th,\mathrm{NT}} = \Delta T_{\mathrm{NT}}/\Delta P_{\mathrm{NT}}$) is determined to be 0.533 $\pm$ 0.008 K/$\mu$W for nanoheater \#1 and  0.485 $\pm$ 0.004  K/$\mu$W for nanoheater \#2. Although the nanoheater TCRs are obtained in the temperature range less than 350 K due to the limit of the heating stage, the linearity measured between $T_{\mathrm{NT}}$ and $P_{\mathrm{NT}}$ signifies that the obtained TCRs are valid for higher temperatures.}

Since the tip side has no sensing component, the nanoheater should measure both the heat transfer rate to the tip ($Q_\mathrm{tip}$) and the sensing region temperature ($T_\mathrm{NT}$) as the tip approaches the nanoheater. To this end, the electrical current is feedback-controlled to compensate tip-induced thermal transport while $T_{\mathrm{NT}}$ is maintained at a set-point value \cite{Jarzembski2018}. \textcolor{red}{For the optimal response time and noise suppression of the nanoheater, an 8$^{\mathrm{th}}$-order low pass filter with 10 Hz cutoff frequency is implemented while the feedback integration gain is set to 20 V/$\Omega$-s. Figure S4 in the Supplemental Material demonstrates the feedback control result of nanoheater \#1 \cite{SI_Jarzembski}. When the temperature set-point is dropped from 481.93 K (or $R_{\mathrm{NT}}=7.12$ $\Omega$) to 479.17 K ($R_{\mathrm{NT}}=7.10$ $\Omega$) and returned back to 481.93 K, $T_{\mathrm{NT}}$ responds to the stepwise set-point changes within a settling time of $\sim$0.5 s and an overshoot temperature of $\sim$0.5 K by changing $P_{\mathrm{NT}}$ by $\sim$3 $\mu$W. The noise-equivalent-temperature (NET) and noise-equivalent-power (NEP) of the feedback-controlled nanoheater sensing region can be determined by conducting a time-based statistical analysis of the $T_{\mathrm{NT}}$ and $P_{\mathrm{NT}}$ traces \cite{bentley1988principles}. When nanoheater \#1 is feedback-contolled at the set-point of 481.93 K and the sampling rate of 500 Hz, its NET and NEP are measured to be 32 mK and 36 nW, respectively. We note that the low pass filter sufficiently eliminates the power noise at 60 Hz, yielding a three times improvement in the NEP when compared with the previous result \cite{Jarzembski2018}. The measured NEP value is in good agreement with the NEP estimated from $R_{th,\mathrm{NT}}$ (i.e., NEP = NET/$R_{th,\mathrm{NT}} \approx$ 60 nW) \cite{Sadat2012a}. The evaluated NEP confirms that the feedback-controlled nanoheater can measure the tip-induced heat transfer rate in the near-contact regime, which can be as small as $\sim$100 nW for the present study.}


\subsection{Experimental procedures}
To consistently describe the thermal and force interactions around contact, we define the tip-nanoheater gap ($d$) as the distance between the mean lines of the tip and nanoheater surface profiles. \textcolor{red}{To avoid any undesired inaccuracy due to surface contamination, both the tip and nanoheater surfaces undergo cleaning procedures outlined in Appendix A. The surface profiles of both the Pt nanoheater sensing region and flattened Si tip are then measured by atomic force microscopy as described in Appendix B.} As shown in Fig. \ref{Fig:gap}(a), the measured surface profiles for both surfaces follow Gaussian distributions, from which the surface roughness is determined to be $R_{p,\textrm{NT}}=5.0\pm0.1$ nm for the nanoheater sensing area and $R_{p,\textrm{tip}}=0.86\pm0.01$ nm for the Si tip within a 98\% confidence interval. As illustrated in Fig. \ref{Fig:gap}(b), the bulk-contact (BC) regime is thus defined as $d\le0$, where the majority of surface asperities are in solid contact. The near-contact (NC) regime is where $d$ is larger than the surface peak heights (i.e., $d>d_\mathrm{AC}$, where $d_\mathrm{AC}\approx R_{p,\textrm{tip}}+R_{p,\textrm{NT}}$) to ensure no contact between surface asperities. The asperity-contact (AC) regime resides between the BC and NC regimes.

\begin{figure} [t!]
  \begin{center}
    \includegraphics[width=0.65\linewidth]{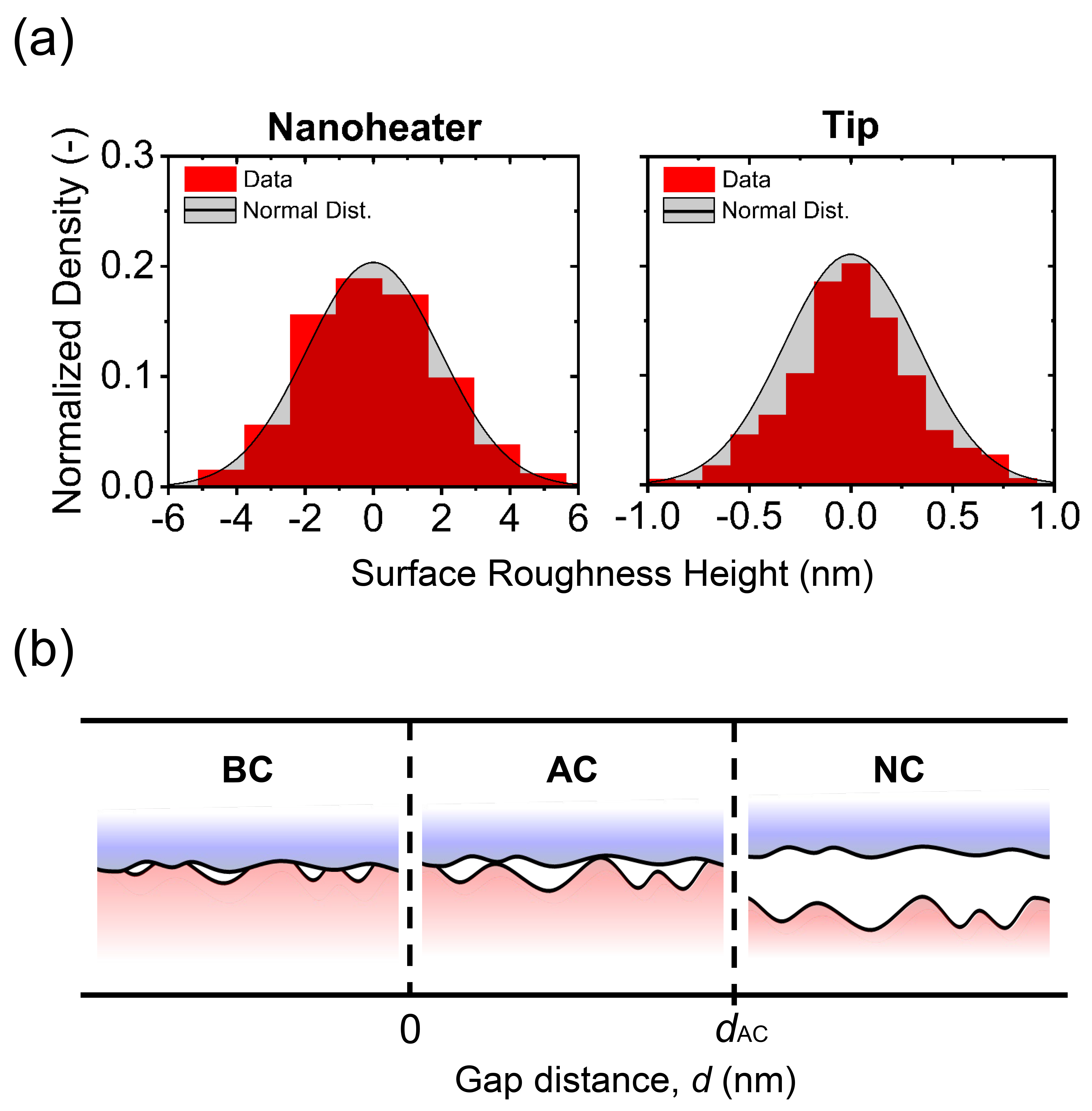}
  \end{center}
  \caption{(a) Measured surface roughness distributions of the nanoheater and tip, which have nominal peak heights of $5.0\pm0.1$ nm and $0.86\pm0.01$ nm, respectively, as measured within a 98\% confidence interval. (b) Illustration of the gap distance ($d$) defined between the mean lines of the flattened Si tip and Pt nanoheater surface profiles showing three regimes defined based on the gap distance, i.e., bulk-contact (BC), asperity-contact (AC), and near-contact (NC).}
  \label{Fig:gap}
\end{figure}

\begin{figure} [t!]
  \begin{center}
    \includegraphics[width=0.55\linewidth]{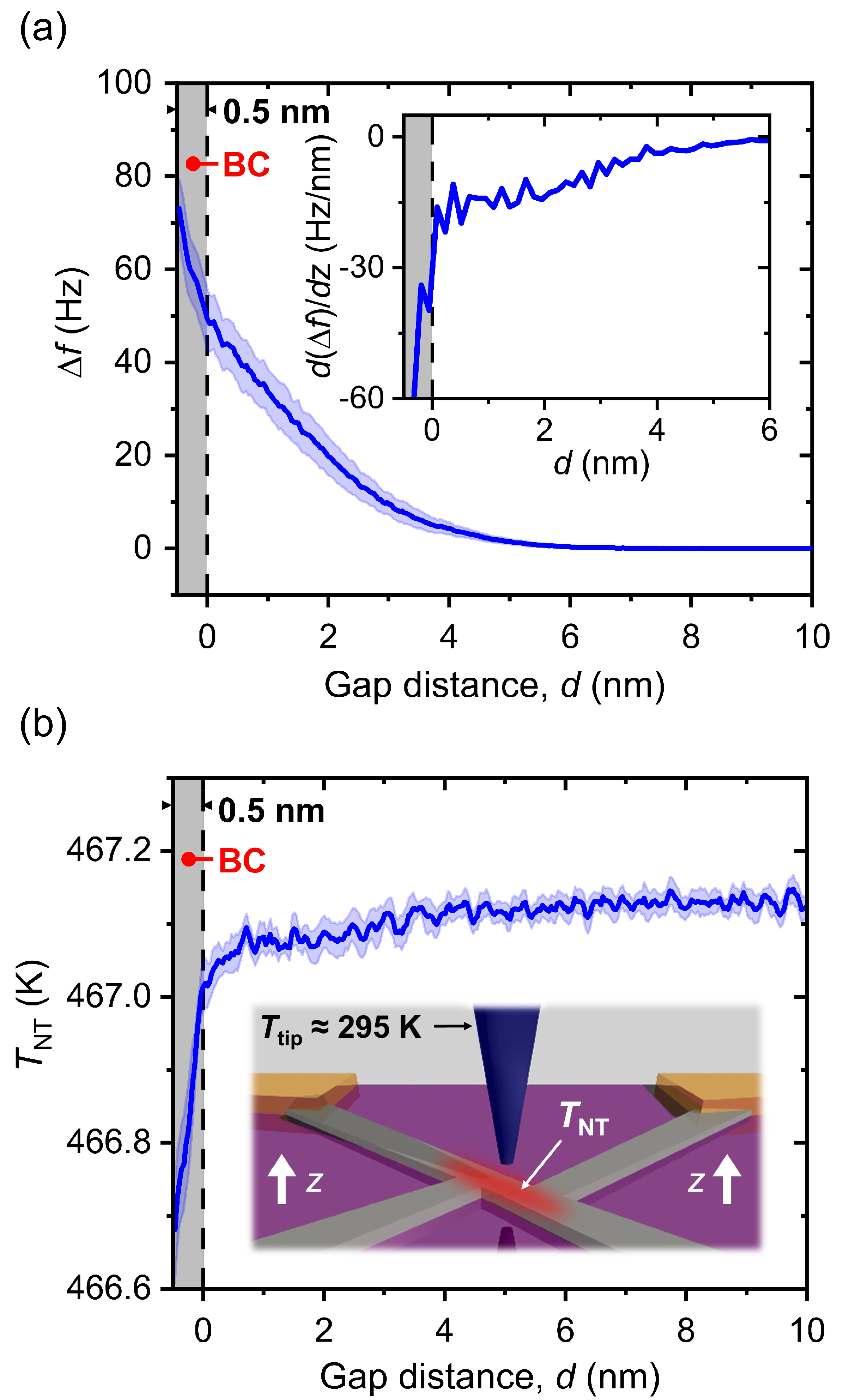}
  \end{center}
  \caption{(a) Measured $\Delta f$ of the QTF, illustrating the monotonic increase of tip-sample lateral forces with decreasing $d$. At BC, this variation becomes more rapid as denoted by its $z$-derivative drop at 0 nm (inset). (b) Feedback-controlled $T_{\mathrm{NT}}$ trace as the nanoheater is approached to the tip at room temperature (i.e., $T_{\mathrm{tip}}$ $\approx$ 295 K). In this figure, the solid lines show the average of 13 measurements whose 95\% confidence interval is denoted by the blue shaded region.}
  \label{Fig:Contact}
\end{figure}

The benefit of combining the feedback-controlled nanoheater and HV-SFM platform is the simultaneous measurement of tip-induced thermal transport and conservative tip-nanoheater lateral force interaction. Figure \ref{Fig:Contact} shows the $\Delta f$ and $T_{\mathrm{NT}}$ traces, respectively, as the Si tip is approached to the sensing region of the feedback-controlled nanoheater. The approaching speed of the sample stage is 0.75 nm/s to provide sufficient time to stabilize $T_{\mathrm{NT}}$ at the set point of 467.13 K within $\pm$50 mK accuracy. While $\Delta f$ monotonically increases as the nanoheater approaches the tip, its \textit{z}-derivative shown in the inset clearly shows a drastic drop when the tip makes bulk contact with the nanoheter surface. The onset of bulk contact can also be confirmed by the $T_{\mathrm{NT}}$ trace, which drops at the mechanically determined BC point. \textcolor{red}{The nanoheater feedback controller, which is optimized to provide the best signal-to-noise ratio in $T_\mathrm{NT}$ and $P_\mathrm{NT}$ measurements, is not fast enough to fully respond to the abrupt increase of conduction heat transfer through the bulk-contacted Si-Pt interfaces. The slow feedback control is also responsible for the slight drift of $T_{\mathrm{NT}}$ by $\sim$0.1 K from the set point when the tip is in the AC regime near the BC point. Nevertheless, this result suggests that the onset of BC can be concurrently determined from mechanical and thermal signals and be used as a reference point for the tip-nanoheater gap distance (i.e., $d=0$).} 

\textcolor{red}{The heat transfer rate from the nanoheater sensing region to the tip ($Q_{\mathrm{tip}}$) can be measured by monitoring $P_{\mathrm{NT}}$ that changes to maintain $T_{\mathrm{NT}}$ at a set-point while the tip approaches the nanoheater sensing region. However, it should be noted that $T_{\mathrm{NT}}$ is the averaged temperature of the nanoheater sensing region. Although $T_{\mathrm{NT}}$ is maintained constant under feedback control, tip-induced local cooling perturbs a temperature distribution of the nanoheater to cause heat conduction from the Joule-heated electrical leads to the sensing region ($Q_\mathrm{lead}$). The tip-based heat transfer rate is thus determined by $Q_{\mathrm{tip}} = \Delta P_\mathrm{NT}+Q_\mathrm{lead}$. Based on the effective thermal network analysis, $Q_{\mathrm{lead}}$ can be modelled as $Q_{\mathrm{lead}}=2\mathcal{L}_{0} T_\mathrm{NT} L_\mathrm{NT}(T_{\mathrm{NT}} \ - \ T_\mathrm{\infty})\Delta P_{\mathrm{H}}/\left(L_\mathrm{lead} R_{\mathrm{NT}}P_{\mathrm{H},0}\right)$,
where $\mathcal{L}_{0}=2.44\times10^-8\ \mathrm{W \Omega/K^{2}}$ is the Lorentz number, $L_\mathrm{NT}$ is the length of the sensing region, $L_\mathrm{lead}$ is the effective length of the lead from the inner electrode to the lead hotspot, and $\Delta P_{\mathrm{H}}$ is the difference of power dissipation in the lead with ($P_{\mathrm{H,1}}$) and without ($P_{\mathrm{H,0}}$) tip-induced cooling. The derivation of $Q_\mathrm{lead}$ and details of notation are provided in Section I.D of the Supplemental Material \cite{SI_Jarzembski}. When $T_{\mathrm{NT}}$ is set to 467.13 K, $Q_\mathrm{lead}$ is estimated to be $\sim$4.2\% of the measured $\Delta P_\mathrm{NT}$. The experimental thermal conductance is then defined as $G_{\mathrm{exp}} = Q_{\mathrm{tip}}/\Delta T$, where $\Delta T = T_\mathrm{NT}-T_\mathrm{tip}$ is the temperature difference between the nanoheater sensing region and the tip apex. Since $T_\mathrm{tip}$ is not directly measurable, it is assumed to be the same as the tip base temperature at $T_{\mathrm{tip}}=295 \pm 0.05 \ \mathrm{K}$. Section I.E of the Supplemental Material supports this assumption by numerically calculating the thermal conductance of the tip, which is two orders of magnitude greater than the BC thermal conductance \cite{SI_Jarzembski}.}  

\subsection{Measurement of $G_\mathrm{exp}$}

\begin{figure} [t!]
  \begin{center}
    \includegraphics[width=0.55\linewidth]{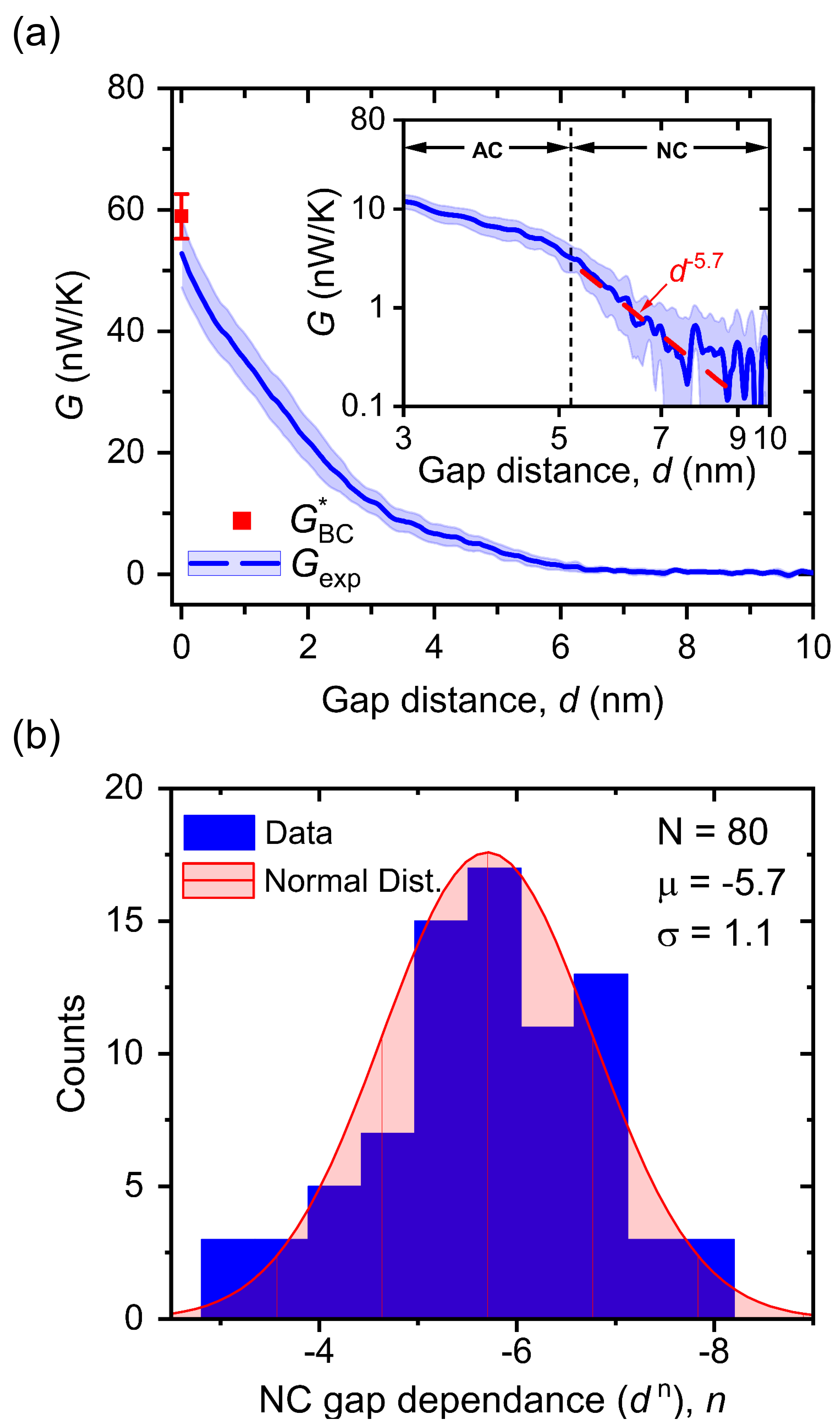}
  \end{center}
  \caption{(a) Measured thermal conductance (\textit{G}$_{\mathrm{exp}}$) as \textit{d} is reduced to BC. Measured thermal conductance replotted on a log-log scale in the inset to show the gap dependence in the NC regime. In this figure, the solid lines show the average of 13 measurements whose 95\% confidence interval is denoted by the blue shaded region. (b) The distribution of the NC gap dependence (\textit{d}$^{n}$) from 80 measurements, demonstrating its repeatability at \textit{n} $= $ -5.7 $\pm$ 1.1.}
  \label{Fig:Exp}
\end{figure}

In Fig. \ref{Fig:Exp}(a), the experimental thermal conductance $G_\mathrm{exp}$ shows a monotonically increasing trend as the gap distance $d$ decreases, approaching the adjusted BC thermal conductance ($G^{*}_{\mathrm{BC}}$) of 58.9 $\pm$ 3.7 nW/K. It should be noted that $G^{*}_{\mathrm{BC}}$ was separately measured using a cantilever probe and adjusted by considering the difference in tip apex geometry: see Appendix C for more details. When $G_\mathrm{exp}$ is re-plotted on a log-log scale as shown in the inset of Fig. \ref{Fig:Exp}(a), the $G_{\mathrm{exp}}$ curve displays different gap dependencies between the NC regime and the AC regime. In the NC regime, $G_{\mathrm{exp}}$ increases by an order of magnitude from the nanoheater noise threshold (i.e., 0.4 nW/K) following a $d^{-5.7}$ power law. In order to confirm the reproducibility of the observed NC gap dependence ($d^{n}$), we repeated the measurement 80 times using three different tip-nanoheater sets and extracted $n$ values. Each data fits well with the $d^n$ function in the NC regime as quantified by an average R$^2$-value of 0.87. Figure \ref{Fig:Exp}(b) shows the histogram of $n$ values, which is well represented by a Gaussian distribution to yield $n=-5.7\pm1.1$. On the other hand, the AC regime exhibits a smaller gap dependence than $d^{-5.7}$ due to the increasing contribution of conduction heat transfer through asperity contacts. In the AC regime, $G_{\mathrm{exp}}$ increases by another order of magnitude approaching $G^{*}_{\mathrm{BC}}$ at $d = 0$. 

\section{Theoretical Modeling}
The experimental thermal conductance $G_{\mathrm{exp}}$ conveys a combination of non-contact and contact heat transfer mechanisms that are complicated to model. The gap dependence of $G_{\mathrm{exp}}$ in the NC regime (i.e., $d^{-5.7\pm1.1}$) is much steeper than what has been predicted with the existing NFRHT models for the tip-plane configuration \cite{Rousseau2009a,Kim2015a,Edalatpour2016,Jarzembski2017}, suggesting that NFRHT may not be the dominant heat transfer mechanism in the NC regime. To elucidate the physics underlying the measured thermal conductance, \textcolor{red}{the theoretical thermal conductance, $G_\mathrm{theory}$, is calculated by considering acoustic phonon transport and NFRHT between tip and nanoheater surfaces that are regenerated based on the measured surface roughness distribution. Since commercial Si microcantilevers are typically n-doped, we consider that the Si tip is n-doped with phosphorus at $1\times10^{18}$ cm$^{-3}$ based on the electrical resistance range provided by the manufacturer. The heat transfer coefficient due to electron tunneling has also been calculated using the framework described in Ref. \cite{Tokunaga2021}, and its contribution has been found to be orders of magnitude smaller than the other heat transfer mechanisms. Throughout the following discussion, the subscripts $\mathrm{L}$ and $\mathrm{R}$ refer to the left and right regions that are respectively made of Pt and Si. The temperatures of the left and right regions are fixed at $T_\mathrm{L}$ = 470 K and $T_\mathrm{R}$ = 300 K.}

\begin{figure} [t!]
  \centering
    \includegraphics[width=0.55\linewidth]{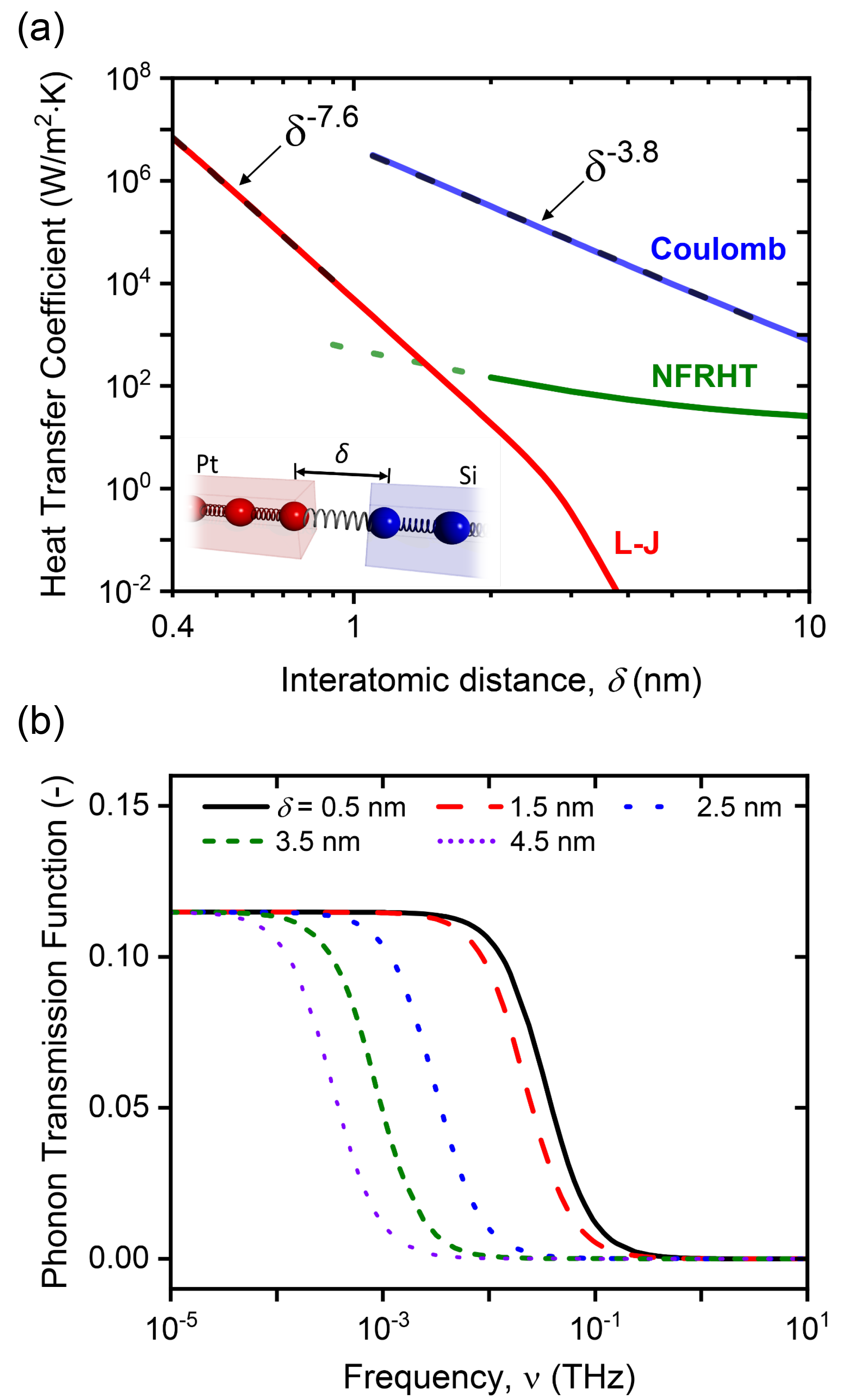}
  \caption{ (a) Theoretical heat transfer coefficients for varying interatomic distance (${\delta}$) computed by fluctuational electrodynamics for NFRHT and the AGF method, where individual AGF contributions are given by the Lennard-Jones (L-J) and Coulomb forces. The possible inaccuracy of NFRHT predictions for ${\delta} < 2$ nm is emphasized via a dotted line. (b) Phonon transmission function through the vacuum space calculated with the AGF method.}
  \label{Fig:Theory}
\end{figure}

\textcolor{red}{\subsection{Heat transfer coefficient contributions}}
Heat transfer due to acoustic phonon transport is calculated via the AGF method \cite{Sadasivam2014} applied to a one-dimensional (1D) Si-Pt atomic chain that has an interatomic vacuum gap distance $\delta$: See the inset of Fig. \ref{Fig:Theory}(a) for the schematic of the 1D atomic chain. \textcolor{red}{The heat flux due to acoustic phonon transport across the interatomic vacuum distance $\delta$ for the 1D atomic chain is given by \cite{Sadasivam2014} 
\begin{equation}
\label{heat flux}
q_\mathrm{ph}  \ = \ \frac{1}{A_\mathrm{Si-Pt}}\int_{0}^{\infty}d{\omega}\frac{\hbar\omega}{2\pi}{\cal T_\mathrm{ph}}({\omega})[N(\omega, T_\mathrm{L})-N(\omega,T_\mathrm{R})]. 
\end{equation} 
Here, $N = 1/[\mathrm{exp}({\hbar\omega/k_\mathrm{b}T})-1]$ is the Bose-Einstein distribution function, where $\hbar$ is the reduced Planck constant and $k_\mathrm{b}$ is the Boltzmann constant. The effective heat transfer area, $A_\mathrm{Si-Pt}$, is the projected atomic area calculated using an average atomic radius, i.e., $A_\mathrm{Si-Pt}  \ = \ \pi\left(r_\mathrm{Pt} + r_\mathrm{Si}\right)^2/4$, 
where $r_\mathrm{Pt} = 1.77{\times}10^{-10}{\ }\mathrm{m}$ and $r_\mathrm{Si} = 1.11{\times}10^{-10}{\ }\mathrm{m}$ are the atomic radius of Pt and Si, respectively \cite{Clementi1967}. The phonon transmission function, ${\cal T_\mathrm{ph}}$, is derived from the AGF method by modeling Pt and Si as semi-infinite leads separated by a device region. The device region contains atoms of Pt and Si separated by the vacuum gap.} \textcolor{red}{The phonon transmission function is written as
\begin{equation}
\label{transmission function}
{\cal T_\mathrm{ph}}({\omega}) \ = \ \mathrm{Trace}[\Gamma_\mathrm{L}G_\mathrm{d}\Gamma_\mathrm{R}G_\mathrm{d}^\mathrm{\dagger}],
\end{equation} 
where the superscript $\dagger$ denotes conjugate transpose. The escape rate of phonons from the device region to the semi-infinite leads, $\Gamma_\mathrm{L(R)}$, is defined as 
\begin{equation}
\label{escape rate}
\Gamma_\mathrm{L(R)} = i[\Sigma_\mathrm{L(R)}-\Sigma_\mathrm{L(R)}^\mathrm{\dagger}],
\end{equation} 
where $\Sigma_\mathrm{L(R)}$ is the self-energy matrix that can be written as $\Sigma_\mathrm{L(R)} = \tau_\mathrm{L(R)}g_\mathrm{L(R)}\tau_\mathrm{L(R)}^{\dagger}$.
Here, $\tau_\mathrm{L(R)}$ is the coupling matrix connecting the left ($\mathrm{L}$) or right ($\mathrm{R}$) semi-infinite lead with the device region. The coupling matrix is computed via the force constant between the atoms bounding the semi-infinite leads and the device region, and the atomic masses. The atomic masses of Pt and Si are $3.239\times10^{-25}{\,}\mathrm{kg}$ and $4.664\times10^{-26}{\,}\mathrm{kg}$, respectively \cite{Meija2016}. The term $g_\mathrm{L,R}$ is the uncoupled Green's function (also called surface Green's function) derived from the harmonic matrix of the left ($\mathrm{L}$) or right ($\mathrm{R}$) semi-infinite lead. The uncoupled Green's function is computed using the decimation technique described in Ref. \cite{Sadasivam2014}. In Eq. (\ref{transmission function}), the device Green's function ($G_\mathrm{d}$) is given by
\begin{equation}
\label{device green's function}
G_\mathrm{d} \ = \ [\omega^{2}I - H_\mathrm{d} - \Sigma_\mathrm{L} - \Sigma_\mathrm{R}]^{-1},
\end{equation} where $I$ is the identity matrix, and $H_\mathrm{d}$ is the harmonic matrix of the device region.} 

\textcolor{red}{Heat transfer due to acoustic phonon transport is mediated by short-range and long-range forces in the vacuum region.} The overlapping electron cloud repulsive force and van der Waals (vdW) force interactions, modeled by the Lennard-Jones (L--J) potential with empirical parameters for Pt and Si \cite{Webb1996}, and the Coulomb force due to surface charges on the nanoheater sensing region, are considered as interatomic forces that virtually connect the Pt and Si atomic chains. \textcolor{red}{Details regarding calculation of force constants acting in the vacuum space are provided in Appendix D.} \textcolor{red}{The force constants are then included in the harmonic matrix $H_\mathrm{d}$ following the procedure described in Appendix E.} \textcolor{red}{The heat transfer coefficient is finally obtained by dividing the heat flux by the temperature difference, i.e., $h_\mathrm{ph}=q_\mathrm{ph}/\Delta T$. Five atoms of Pt and five atoms of Si in the device region are sufficient to obtain stable and converged results.} Note that \textcolor{red}{although the 1D AGF calculation does not capture the angle dependence of phonon propagation and force interactions with neighboring atoms, a previous work \cite{Chiloyan2015b} demonstrated that phonon transport becomes quasi 1D in the NC regime. This has also been verified by comparing 1D AGF results against three-dimensional lattice dynamics results \cite{Alkurdi2020}: More discussion on the verification of the 1D AGF method for calculating interfacial and near-contact acoustic phonon transport is provided in Section II of the Supplemental Material \cite{SI_Jarzembski}.} 

\textcolor{red}{The heat transfer coefficient due to NFRHT is calculated using fluctuational electrodynamics \cite{Rytov1989}, where Pt and Si are modeled as two semi-infinite planes \cite{Polder1971,Tokunaga2021}. The dielectric functions are taken from Ref. \cite{Djurisic1997} for Pt and Refs. \cite{Fu2006,Basu2010a} for n-doped Si. The radiative heat transfer coefficient is calculated down to an interatomic vacuum distance $\delta$ of 1 nm. However, since fluctuational electrodynamics is a theory based on the macroscopic Maxwell equations that is unlikely to be valid for such a small distance, the NFRHT results below $\delta=2$ nm are plotted with a dotted curve.}

\textcolor{red}{Figure \ref{Fig:Theory}(a) shows the individual contribution of the L--J and Coulomb forces to the AGF-calculated acoustic phonon heat transfer coefficient for the 1D Si-Pt atomic chain. While the L--J force model drives heat transfer for $\delta$ $<$ 1.1 nm, the Coulomb force becomes a dominant contributor for larger $\delta$ values (i.e., 1.1 nm $<$ $\delta <$ 10 nm). This significant contribution of the Coulomb force arises from surface charges induced by the local voltage bias at the center of the nanoheater sensing region (0.8 V from the ground) when it is Joule-heated at $T_\mathrm{NT}=$ 467.13 K. Here, we assume that the Coulomb force vanishes for $\delta<$ 1.1 nm due to surface charge neutralization between the nanoheater and tip: See also Appendix D. As a result, acoustic phonon heat transfer between Pt and Si surfaces can exceed NFRHT for interatomic distances up to $\delta\sim$ 10 nm. The potential inaccuracy of the NFRHT calculations has no impact on the theoretical thermal conductance since heat transfer is largely dominated by acoustic phonon transport for $\delta <$ 2 nm. The calculated heat transfer coefficients follow power laws of $\delta^{-7.6}$ for the L-J force and $\delta^{-3.8}$ for the Coloumb force, respectively, illustrating that the experimental value of $d^{-5.7}$ could be indicative of acoustic phonon transport. Figure \ref{Fig:Theory}(b) shows the phonon transmission function calculated by the AGF method, where the dominant frequency range of phonon transmission is below $\sim$1.0 THz at $\delta =$ 0.5 nm. The phonon dispersion curves and density of states for bulk Pt and Si \cite{Sun2008, Esfarjani2011} confirm that acoustic phonons are the dominant heat carriers across the vacuum distance in the NC regime.}

\textcolor{red}{\subsection{Comparison between $G_\mathrm{exp}$ and $G_\mathrm{theory}$}}

{\color{red} For fair comparison with $G_{\mathrm{exp}}$, the theoretical thermal conductance ($G_{\mathrm{theory}}$) is calculated by implementing surface features of the flattened Si tip and the Pt nanoheater sensing region in the AGF calculation. To this end, both surfaces are randomly regenerated from the measured surface roughness distributions shown in Fig. \ref{Fig:gap}(a) and discretized into $N$ flat pixels having different gap distances. Once the local heat transfer coefficient $h_i$ is calculated by applying the 1D AGF for each pixel, the thermal conductance is approximated as $G_{\mathrm{theory}}\left(d\right) = \sum_{i=1}^{N}h_{i}\left(\delta_i\right)A_i$, where $A_i$ is the pixel area and $\delta_i$ is the local interatomic distance for each pixel. The minimum value of $\delta_i$ (i.e., contact) is set to 4.68 \AA, which is the average value of the lattice constants for Pt (3.92 \AA) \cite{Feibelman2001} and Si (5.43 \AA) \cite{Esfarjani2011}. The effective heat transfer area, limited by the Si tip surface, is determined based on the SEM image shown in Fig. 1(c). We approximate the Si tip surface as a square whose diagonal length is 240 nm, yielding $A_\mathrm{tip} = \left(240\times10^{-9}/\sqrt{2}\right)^2 =2.88 \times 10^{-14} \ \mathrm{m^2}$. $A_\mathrm{tip}$ is discretized into $N$ sub-surfaces of equal size (i.e., $A_i = A_{\mathrm{tip}}/N)$.
A convergence analysis revealed that $N = 1024$ is sufficient to obtain stable results. In order to develop statistically relevant predictions, $G_{\mathrm{theory}}$ is calculated from 30 regenerated surfaces, which results in 30 sets of $G_\mathrm{theory}(d)$.} 

\begin{figure} [t!]
\centering
    \includegraphics[width=0.55\linewidth]{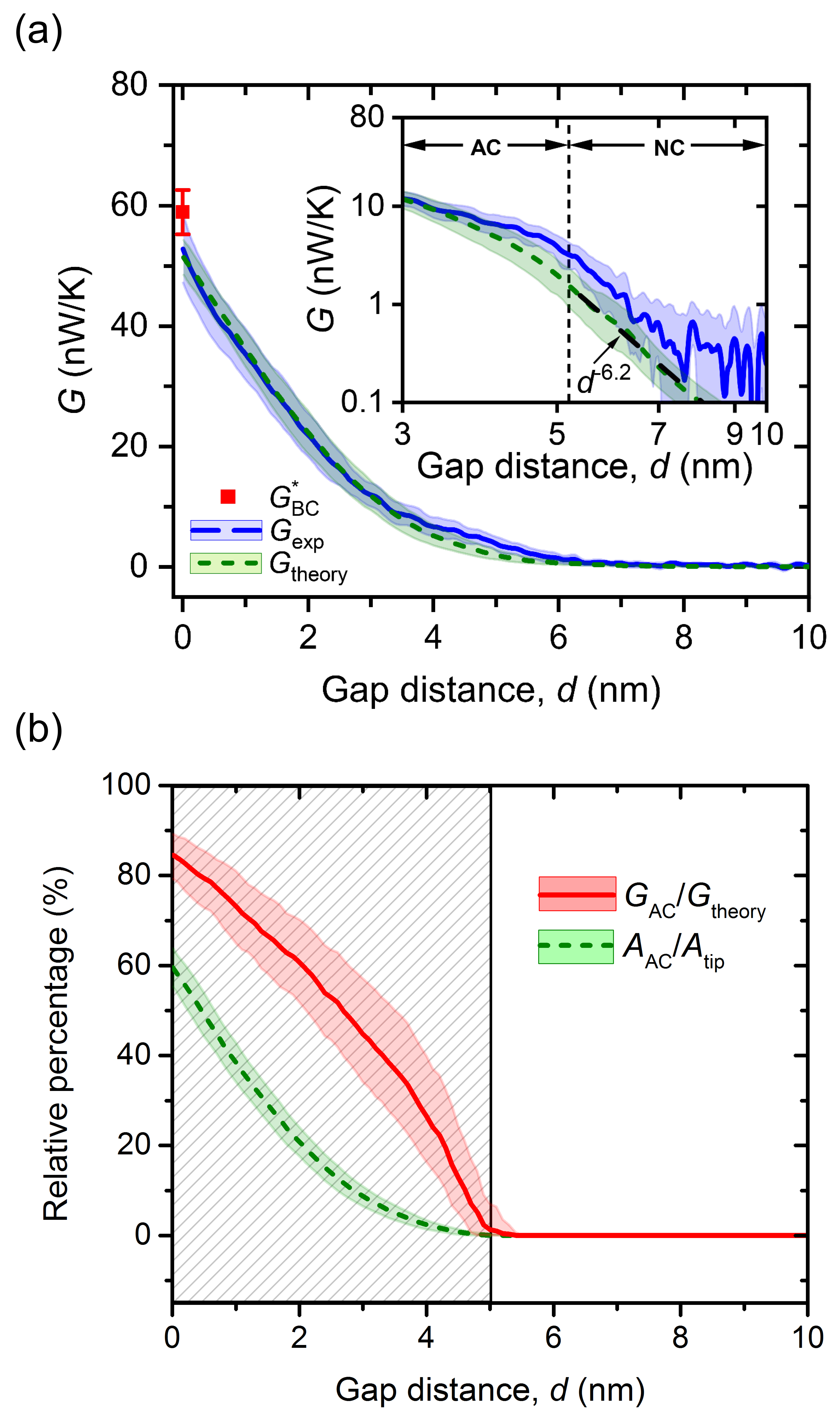}
\caption{(a) Theoretical and experimental thermal conductances on linear-linear and (inset) log-log scales. (b) Calculated AC thermal conductance ratio (\textit{G}$_{\mathrm{AC}}$/\textit{G}$_{\mathrm{theory}}$, where \textit{G}$_{\mathrm{theory}}$ = \textit{G}$_{\mathrm{NC}}$ + \textit{G}$_{\mathrm{AC}}$) and asperity contact area ratio (\textit{A}$_{\mathrm{AC}}$/\textit{A}$_{\mathrm{tip}}$), showing the transition between the NC and AC regimes.}
\label{Fig:ExpTheory}
\end{figure}

Figure \ref{Fig:ExpTheory}(a) compares $G_{\mathrm{theory}}$ and $G_{\mathrm{exp}}$ on linear-linear and log-log scales (inset), demonstrating strong agreement between them for both NC and AC regimes. \textcolor{red}{The bold dashed lines (green color) correspond to the theoretical thermal conductance averaged over 30 regenerated surfaces, while the shaded regions (green color) are produced by calculating $G_\mathrm{theory}^{\mathrm{average}}\pm2\sigma$, where $\sigma$ is the standard deviation for the upper and lower bounds of the surface charge density: See Appendix D. The gap dependence of $G_{\mathrm{theory}}$ in the NC regime is $-6.2\pm0.4$, which is consistent with the measured gap dependence of $d^{-5.7\pm1.1}$ within the uncertainty. In addition, the theoretical thermal conductance at $d=0$ (54.2 nW/K) is in good agreement with the bulk thermal conductance values measured with the QTF probe (52.8 nW/K) and the cantilever probe (58.9 nW/K).} These well-agreed experimental and theoretical results strongly suggest that acoustic phonon transport play a significant role in heat transfer between Si and Pt for both the NC and AC regimes, possibly being the mechanism bridging radiation and conduction heat transfer. Moreover, the AGF method can separate the contributions of phonon transport through the vacuum gap ($G_\mathrm{NC}$) and asperity contacts ($G_{\mathrm{AC}}$) towards the total thermal conductance (i.e., $G_{\mathrm{theory}}=G_\mathrm{NC}+G_{\mathrm{AC}}$). Figure \ref{Fig:ExpTheory}(b) shows the asperity-contact thermal conductance ratio ($G_{\mathrm{AC}}/G_{\mathrm{theory}}$) and the asperity-contact area ratio ($A_{\mathrm{AC}}/A_{\mathrm{tip}}$, where $A_{\mathrm{AC}}$ is the asperity-contact area and $A_{\mathrm{tip}}$ is the total tip area). While both $G_{\mathrm{AC}}/G_{\mathrm{theory}}$ and $A_{\mathrm{AC}}/A_{\mathrm{tip}}$ are $0$\% in the NC regime signifying non-contact heat transfer, they start to increase at $d_{\rm AC}\approx 5.2$ nm due to the onset of asperity contacts. It should be noted that the vertical dashed line drawn in the inset of Fig. \ref{Fig:ExpTheory}(a) denotes $d_{\rm AC}$, which is in agreement with the aforementioned definition of the NC and AC regimes (i.e., $d_{\rm AC}\approx R_{p,\mathrm{tip}}+R_{p,\mathrm{NT}}$) based on the surface roughness. At $d = 0$, $G_{\mathrm{AC}}$ reaches $\sim$80\% of $G_{\mathrm{theory}}$ while $A_{\mathrm{AC}}$ becomes $\sim$60\% of $A_{\mathrm{tip}}$, theoretically supporting the smooth transition from near-contact to bulk-contact heat transfer by means of acoustic phonon transport. 

\begin{figure} [t!]
\centering
    \includegraphics[width=0.5\linewidth]{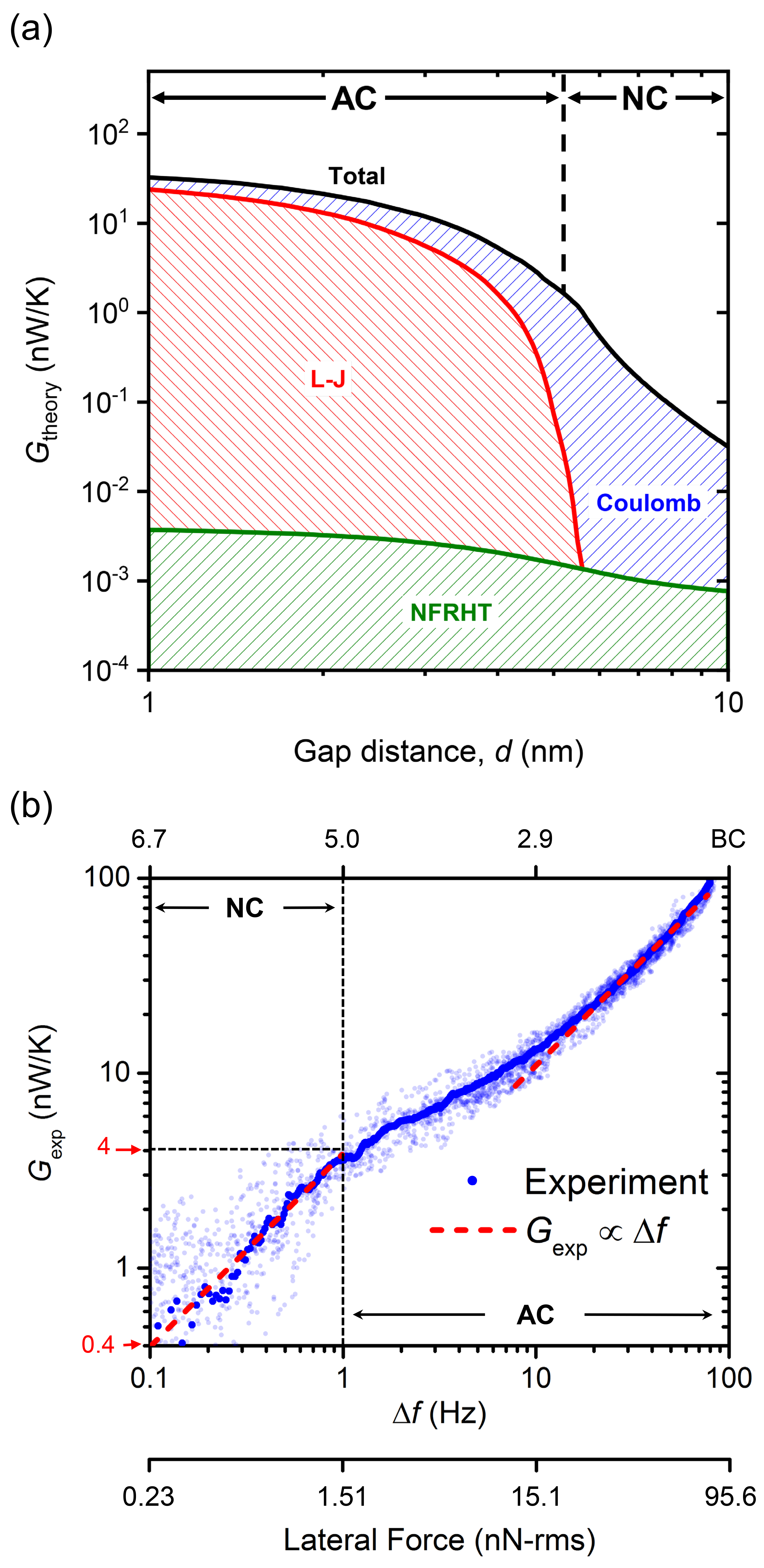}
\caption{(a) AGF force model contributions to \textit{G}$_{\mathrm{theory}}$ are compared with NFRHT, which is provided as a reference. (b) $G_{\mathrm{exp}}$ correlation with the simultaneously acquired $\Delta$\textit{f}, where $\Delta$\textit{f} is related to the tip-sample lateral force.}
\label{Fig:Force}
\end{figure}

Since acoustic phonon transport across a vacuum gap is mediated by interatomic force interactions between Pt and Si atoms, the impact of each force contributing to $G_{\mathrm{theory}}$ is calculated with the AGF method and shown in Fig. \ref{Fig:Force}(a). Here, NFRHT is also included for comparison. In the AC regime, the short-range force interactions, such as the repulsive force due to overlapping electron clouds and the vdW force, dominate thermal transport. However, the Coulomb force becomes a dominant contributor in the NC regime, allowing acoustic phonon transport to exceed NFRHT by up to three orders of magnitude. When considering the origin of the Coulomb force, our calculation suggests the possibility of manipulating heat transfer in the NC regime with external force stimuli \cite{Pendry2016}. Figure \ref{Fig:Force}(b) further demonstrates the strong correlation between tip-surface heat transfer and force interactions by comparing $G_{\mathrm{exp}}$ to the simultaneously acquired $\Delta f$ of the QTF.  As the contact regime transitions from AC to BC, $G_{\mathrm{exp}}$ becomes linearly proportional to $\Delta f$ as indicated by the red dashed line in the top right corner of Fig. \ref{Fig:Force}(b). \textcolor{red}{Since the lateral force gradient exerted on the tip can be first-order approximated as $\partial F_{x}/\partial z \approx 2k_{\mathrm{eff}} \Delta f/f_0$ \cite{Castellanos-Gomez2011}, the lateral force at different gaps can be calculated by integrating the equation over the interval $d \leq z \leq \infty$. 
The calculated lateral force is denoted in Fig. \ref{Fig:Force}(b) along with $\Delta f$. Since the lateral force can be related with the normal force by the nanoscale friction law \cite{Mo2009}, the observed linear proportionality in the AC-to-BC transition indicates a strong correlation between the normal contact force and the interfacial thermal conductance. Interestingly, a similar linear proportionality between $G_{\mathrm{exp}}$ and $\Delta f$ is observed in the NC regime as indicated by the red dashed line in the bottom left corner, implying that Pt and Si atoms are still connected by interatomic forces to allow non-contact acoustic phonon transport. We also experimentally demonstrate that the tip-sample lateral force increases with the increasing electrical current supplied to the nanoheater in the NC regime (Section III of the Supplemental Material \cite{SI_Jarzembski}), which we believe should increase the thermal conductance. However, the manipulation of the near-contact thermal conductance by external force stimuli was not measured due to limitations in the current nanoheater design, which remains a future research.}

\section{Conclusions}
We have conducted experiments of thermal transport between a flattened Si tip and feedback-controlled Pt nanoheater in a high-vacuum shear force microscope (HV-SFM) as the tip is positioned in the near-contact (NC), asperity-contact (AC), and bulk-contact (BC) regimes with the nanoheater surface. The obtained experimental results clearly show that heat transfer in the NC regime is much greater than NFRHT with a stronger gap dependence. Comparison of the experimental data with heat transfer calculations based on the atomistic Green's function method and fluctuational electrodynamics frameworks provides evidence that acoustic phonons can be transported not only through asperity contacts but also across nanoscale vacuum spaces due to force interactions between terminating atoms separated by vacuum. This finding sheds light on the possibility of engineering interfacial thermal transport using external force stimuli, which can impact the development of next-generation thermal management technologies.

\section*{{\large Acknowledgements}}
This work was supported by the National Science Foundation (CBET-1605584) and the Nano Material Technology Development Program (2015M3A7B7045518) through the National Research Foundation of Korea (NRF). A.J. acknowledges financial supports from the University of Utah's Sid Green Fellowship and the National Science Foundation Graduate Research Fellowship (No. 2016213209). T.T. acknowledges support from the Yamada Science Foundation and the Fujikura Foundation. C.S. acknowledges support from the National Science Foundation Graduate Research Fellowship (No. 2017249785). 
The support and resources from the Center for High Performance Computing at the University of Utah are gratefully acknowledged. 3D schematics were generated using the open-source software POV-ray (www.povray.org). We also thank Prof. Takuma Shiga at the University of Tokyo for fruitful discussions.


\textcolor{red}{\section*{{\large Appendix}}}
\textcolor{red}{\subsection{Surface cleaning procedures}}
\textcolor{red}{To ensure that the interacting surfaces are free from contamination prior to experiments, a routine surface inspection and cleaning protocol was established for the QTF probes and nanoheaters. First, several QTF probes and nanoheaters are inspected using scanning electron microscopy (SEM) to select the ones with no major debris around the sensing areas: see Figs. \ref{Fig:Schematic}(c) and (d). After initial sonication cleaning with acetone, they are placed in a deep ultraviolet (UV) ozone cleaner (Novascan, PSD-UV4) to remove organic contamination using UV light at 185 and 254 nm in wavelength \cite{Tsao2007}. The UV-ozone cleaner is set to generate ozone for 2 hours. After UV treatment, the QTF probe and the nanoheater are promptly mounted to the HV-SFM, which is evacuated to high vacuum to minimize undesired exposure to the ambient before experiments \cite{Cui2017a}. Moreover, all experiments were carried out at high temperature (i.e., 467 K), which inherently removes weakly bonded contaminants from the surface.}

\textcolor{red}{\subsection{Surface roughness characterization}}
\textcolor{red}{In order to secure a sufficient heat transfer area, the Si tip was flattened by long-line scanning on the nanoheater substrate (i.e., a 500-nm thick silicon nitride film on top of a silicon substrate) at a contact force of $\sim$10 nN. Once the flattened tip was attached to the QTF, the tip was long-line scanned on the nanoheater substrate again in asperity contact mode for the fine adjustment of surface parallelism. Since the surface profiles play a pivotal role in determining different contact regimes, we measured the surface roughness distributions of both the nanoheater sensing region and flattened Si tip. The nanoheater surface profile was obtained by soft-contact AFM imaging, as marked by a yellow rectangle in Fig. \ref{Fig:Schematic}(d). A small contact force ($\lesssim$ 3 nN) during the AFM imaging results in a contact diameter of $\sim$ 8 nm as estimated by the Hertzian model \cite{Derjaguin1975}. 
A surface roughness histogram of the nanoheater sensing region from the obtained AFM image shows a Gaussian distribution centered at 0 nm with a standard deviation of $\sigma =$ 1.96 nm: see Fig. \ref{Fig:gap}(a). The equivalent surface peak height is $R_{p,\textrm{NT}}=5.0\pm0.1$ nm within a 98\% confidence interval. 
The surface profile of a flattened Si tip was determined by tapping-mode topographic imaging of a calibration sample consisting of sharp pyramids (K-TEK Nanotechnology, TGT1), whose apex radii are nominally 17 nm.  Since the sample pyramids are much sharper than the flattened Si tip, the resulting convoluted tip-sample AFM image provides the surface roughness profile of the flattened tip. 
A surface roughness histogram of the flattened tip area from the convoluted AFM image displays a Gaussian distribution centered at 0 nm with a standard deviation of $\sigma =$ 0.33 nm, as shown in Fig. \ref{Fig:gap}(a). The equivalent surface peak height is $R_{p,\textrm{tip}}=0.86\pm0.01$ nm within a 98\% confidence interval.} 

\textcolor{red}{\subsection{Measurement of bulk-contact thermal conductance}}
\textcolor{red}{To fully understand the transition from NC to BC thermal transport, the tip should approach the nanoheater sensing region to form BC while avoiding damage to the tip and nanoheater. However, the high vertical rigidity of the QTF can easily damage both the tip and nanoheater when the tip is further pushed once BC is made. To address this challenge, we conducted BC measurements separately by using a cantilever in high vacuum (see Fig. S7(a) in the Supplemental Material \cite{SI_Jarzembski}). It should be noted that our HV-SFM also has a regular AFM head for cantilever-based operations. The cantilever used in the BC experiment is the same model (Bruker, FMV-A) as mounted to the QTFs for the NC and AC measurements. The cantilever deflection is detected by an optical fiber interferometer aligned with the cantilever's backside \cite{Rugar1989}. Nanoheater \#2 was used for the BC measurement (topography shown in  Fig. S7(b)), whose sensing area is 330 nm $\times$ 375 nm. After AFM topographic imaging with soft-contact mode ($F_{\mathrm{z}}\lesssim$ 3 nN), the force spectroscopy measurement was conducted by approaching the tip to the nanoheater sensing region until they make hard contact ($F_{\mathrm{z}}\gtrsim$ 15 nN). An SEM image of the Si tip after the force spectroscopy is shown in Fig. S7(c). We believe that bulk contact is made at the flattest portion of the tip apex to form the contact diameter of 215$\pm$25 nm.}

\begin{figure}[t!]
\centering
\includegraphics[width=0.75\linewidth]{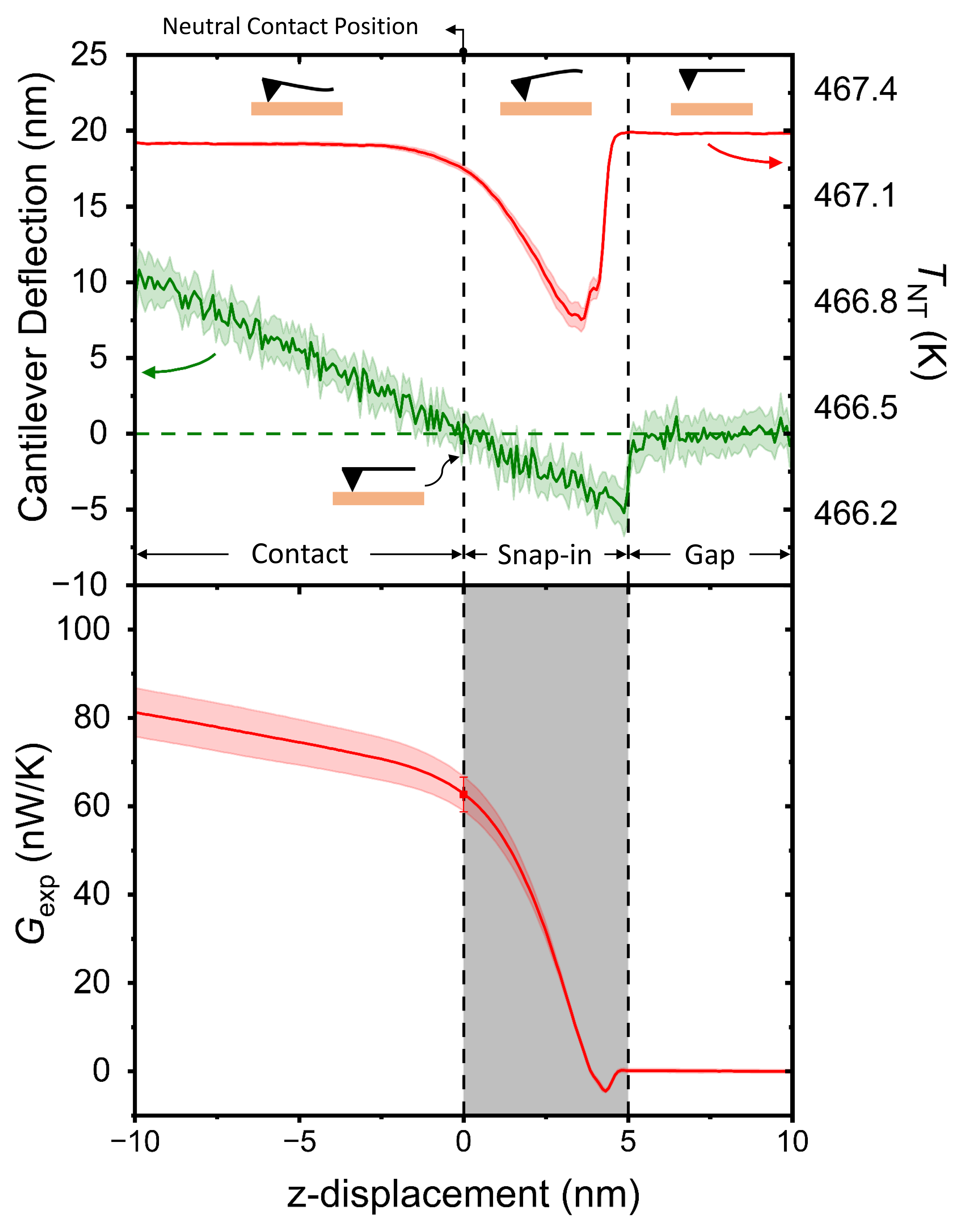}
\caption{Measurement of the bulk-contact thermal conductance (\textit{G}$\bm{_{\mathrm{BC}}}$) for the Si-Pt system. Cantilever deflection and feedback-controlled temperature signal (top), and the measured value of \textit{G}$_{\mathrm{exp}}$ (bottom) are plotted as a function of the $z-$displacement. Here, \textit{G}$_{\mathrm{BC}}$ for the Si-Pt system at 467 K was measured to be 62.7 $\pm$ 3.9 nW/K extracted from a cantilever deflection of 0 nm (i.e., normal force of $\sim$15 nN). The shaded regions show the 95\% confidence interval of 13 measurements, while the solid line represents their average value.
\label{Fig:BC_Exp}}
\end{figure}

\textcolor{red}{For both the cantilever- and QTF-based experiments, the nanoheater current ($I_{\mathrm{S}}$) is feedback-controlled while the voltage drop across the sensing region ($V_{\mathrm{NT}}$) is measured for real-time monitoring of $T_{\mathrm{NT}}$. The only difference in the cantilever-based measurements is that after snap-in-contact is made, the cantilever is further pushed to achieve BC between the tip and nanoheater. Figure \ref{Fig:BC_Exp} presents the cantilever deflection and $T_{\mathrm{NT}}$ signals as a function of the tip position. At a distance of 5 nm, the cantilever snaps into contact as denoted by the sudden drop of the cantilever deflection and $T_{\mathrm{NT}}$. As the tip is continuously pushed toward the nanoheater, the cantilever returns to its neutral position, which is referred to as the zero \textit{z}-displacement position \cite{Kim2015a}. In the negative displacement regime, the sample pushes the cantilever to bend backwards such that bulk contact is made with a sufficient contact force. Meanwhile, the feedback controller settles $T_{\mathrm{NT}}$ to the set-point by increasing the heating power. Figure \ref{Fig:BC_Exp} also shows the corresponding value of $G_{\mathrm{exp}}$, which remains zero in the gap region, rapidly increases at snap-in, and gradually increases as the cantilever is further pushed. The gradual increase of the thermal conductance is indicative of the onset of bulk contact and is attributed to the pressure dependence of the interfacial thermal resistance \cite{Gotsmann2013}. From this measurement, the BC thermal conductance ($G_{\mathrm{BC}}$) between the Si tip and Pt nanoheater is determined to be 62.7 $\pm$ 3.9 nW/K at $z=0$ nm. For proper comparison with the QTF-based experiments, the obtained  $G_{\mathrm{BC}}$ is adjusted by considering the different effective contact area. The adjusted thermal conductance, $G^{*}_{\mathrm{BC}}$, is estimated to be 58.9 $\pm$ 3.7 nW/K and is shown in  Figs. \ref{Fig:Exp}(a) and \ref{Fig:ExpTheory}(a).}

\textcolor{red}{\subsection{Calculation of force constants for the AGF method}}
\textcolor{red}{The short-range electron cloud interaction and van der Waals force are modeled via the L--J potential, while the long-range electrostatic surface charge interaction is modeled with the Coloumb force \cite{Shockley1948}. The L--J force constant is given by:}
\textcolor{red}{\begin{equation}
\label{L-J force constant}
k_\mathrm{L-J}  \ = \ \left| 24{\varepsilon}\left[26\left(\frac{{\sigma}^{12}}{{\delta}^{14}}\right) - 7\left(\frac{{\sigma}^{6}}{{\delta}^{8}}\right)\right]\right|
\end{equation}} 
\textcolor{red}{where $\varepsilon = 4.80{\times}10^{-20}{\ }\mathrm{J}$ and $\sigma = 1.84{\times}10^{-10}{\ }\mathrm{m}$ for the interaction between Pt and Si atoms (Pt-Si)\cite{Webb1996}}. \textcolor{red}{The interatomic force constant of Pt, $k_\mathrm{Pt} = 6.31 \ \mathrm{N/m}$, is obtained from Eq. (\ref{L-J force constant}) using $\varepsilon = 1.09{\times}10^{-19}{\ }\mathrm{J}$ and $\sigma = 2.54{\times}10^{-10}{\ }\mathrm{m}$ for Pt-Pt interaction \cite{Webb1996}, whereas $k_\mathrm{Si} = 6.16 \ \mathrm{N/m}$ is taken from Ref. \cite{Ezzahri2014} for Si-Si interaction. Although the tip used in the experiments is made of n-doped Si (phosphorus-doped at $\rm{1\times10^{18}~{cm^{-3}}}$), it is treated as intrinsic Si in the AGF calculations because the elastic constant of n-doped Si is nearly the same as that of intrinsic Si for doping levels up to $\rm{8.5{\times}10^{18}~{cm^{-3}}}$.\cite{Ono2000}}

\textcolor{red}{The Coulomb force is mediated by surface charges. The Pt nanoheater sensing region has negative surface charges due to the applied bias voltage $V_\mathrm{bias}$ of 0.8 V as measured from the ground. By conceptualizing the Si tip as a floating ground, positive image charges are induced at the tip apex. Surface charges of opposite signs are the source of the Coulomb force. The Coulomb force constant due to surface charges is given by \cite{Terris1989}:
\begin{equation}\label{Coulomb force constant explicit}
k_\mathrm{Coulomb}  \ = \left| \frac{Q_\mathrm{s}}{4\pi\varepsilon_\mathrm{0}}\left(\frac{2Q_s}{\delta^3}+\frac{3{\varepsilon_0}{V_{\mathrm{bias}}}{A_\mathrm{tip}}}{\delta^4}\right)\cdot\frac{A_\mathrm{Si-Pt}}{A_\mathrm{tip}} \right|
\end{equation} 
where $\varepsilon_{0}$ is the permittivity of free space, and $Q_\mathrm{s}$ is the surface charge (= $\sigma_{\mathrm{s}} A_{\mathrm{tip}}$, where $\sigma_{\mathrm{s}}$ is the surface charge density).}
\textcolor{red}{Prediction of $\sigma_{\mathrm{s}}$ is challenging as it depends on the material properties, bias voltage, temperature, and gap distance. To constrain these parameters for our experimental condition, we extract $\sigma_{\mathrm{s}}$ from the the gap-dependent $\Delta f$ signal of the QTF at $d$ = 6 nm, where $\Delta f$ is independently measured from the nanoheater signals and contains the effects of those parameters onto the tip-sample force. It should be noted that the Coulomb force is expected to be dominant at $d$ = 6 nm, which is in the NC regime.}

\textcolor{red}{At $d$ = 6 nm, the lateral force is calculated to be 0.42 nN-rms. In addition, we assume $\mu$ = 0.0005 as the near-contact friction coefficient that correlates the lateral force with the normal force. This value is within the acceptable range for the Si-Pt system with a nanoscale gap because experimental measurements of the contact $\mu$ value ranges from $\sim$0.1 to $\sim$0.01 depending on the contacting area \cite{Bhushan2007}. Furthermore, molecular dynamics simulations for lubricated atomically flat surfaces in contact predicted $\mu\approx0.001$ \cite{Gao2004}. Using $\mu=0.0005$, $\sigma_{\mathrm{s}}$ is estimated to be $8\times10^{-4}\ \mathrm{C/m^2}$ using the Coulomb force equation \cite{Tokunaga2021}. We establish a confidence interval for the near-contact $\mu$ ranging from 0.001 to 0.0004, which corresponds to a surface charge density range of $6\times10^{-4}$ to $1\times10^{-3}\ \mathrm{C/m^2}$. These surface charge density values are in the reasonable range when compared with previous works \cite{Yang2007b,Johann2009,ElKhoury2016,Klausen2016} and used to determine the theoretical uncertainties in conjunction with the surface roughness distributions as shown in Figs. \ref{Fig:ExpTheory}(a) and (b).}
\textcolor{red}{It should be noted that at a  small vacuum distance before contact, the Coulomb force vanishes due to charge neutralization \cite{Behrens2001}. The present work treats $\sigma_{\mathrm{s}}$ as a constant value that vanishes at a specific cutoff gap distance \cite{Butt1991} due to the difficulty of describing its gap-dependence. The cutoff distance is determined at the onset of electron tunneling across the vacuum gap, which is defined at an interatomic distance ${\delta_c}{\,}\mathrm{\approx 10{\,}{\mbox{\AA}}}$ \cite{Anselmetti1994,McCarty2008}.}

\textcolor{red}{\subsection{Incorporation of force constants into the harmonix matrix}}

\textcolor{red}{The force constants of the vacuum region due to the Lennard-Jones potential ($k_{\mathrm{L-J}}$) and Coulomb interactions ($k_{\mathrm{Coulomb}}$) are incorporated into the harmonic matrix $H_{\mathrm{d}}$ in Eq. (\ref{device green's function}) as follows \cite{Sadasivam2014}:} 
\textcolor{red}{\begin{gather}
 \boldsymbol{H}_{\mathrm{d}} \ = \
 \begin{bmatrix}
 \ddots & \vdots & \vdots & \vdots & \vdots & \iddots \\
 \cdots & H^{4,4}_{\mathrm{d}} & -k_{\mathrm{Pt}}/m_{\mathrm{Pt}} & {0} & {0} & \cdots \\
  \cdots & -k_{\mathrm{Pt}}/m_{\mathrm{Pt}} & H^{5,5}_{\mathrm{d}} & -(k_{\mathrm{L-J}} + k_{\mathrm{Coulomb}})/m_{\mathrm{Si}} & {0} & \cdots \\
   \cdots & {0} & -(k_{\mathrm{L-J}} + k_{\mathrm{Coulomb}})/m_{\mathrm{Pt}} & H^{6,6}_{\mathrm{d}} & -k_{\mathrm{Si}}/m_{\mathrm{Si}} & \cdots \\
 \cdots & {0} & {0} & -k_{\mathrm{Si}}/m_{\mathrm{Si}} & H^{7,7}_{\mathrm{d}} & \cdots \\
 \iddots & \vdots & \vdots & \vdots & \vdots & \ddots
 \end{bmatrix}
 \label{Eq:harmonic_matrix_devie}
\end{gather}}
\textcolor{red}{The components of the harmonic matrix are described as $H^{j,k}_{\mathrm{d}}$ $(j,k = 1,2,...,10)$. The force constants of the Pt and Si atoms are given by $k_{\mathrm{Pt}}$ and $k_{\mathrm{Si}}$, and their atomic weight are denoted as $m_{\mathrm{Pt}}$ and $m_{\mathrm{Si}}$, respectively. The device region includes a total of 10 atoms (5 Si atoms for the left side and 5 Pt atoms for the right side), thus resulting in a 10 $\times$ 10 harmonic matrix. The diagonal components of $H_{\mathrm{d}}$ are calculated by} 
\textcolor{red}{\begin{equation}
\label{eqn:HdCalculation}
     H^{j,j}_{\mathrm{d}} \ = \ -\left(H^{j,j-1}_{\mathrm{d}} + H^{j,j+1}_{\mathrm{d}}\right)
\end{equation}}

%% file: PRB_SI.tex
\begin{center}
\textbf{{\small Supplemental Material} \\ \vspace{10pt} \large Role of Acoustic Phonon Transport in Near- to Asperity-Contact Heat Transfer}
\end{center}

\section*{I. Experiments}
\subsection*{A. Z-piezo calibration of the HV-SFM sample stage} 
\begin{figure}[p!]
\centering
\includegraphics[width=0.6\linewidth]{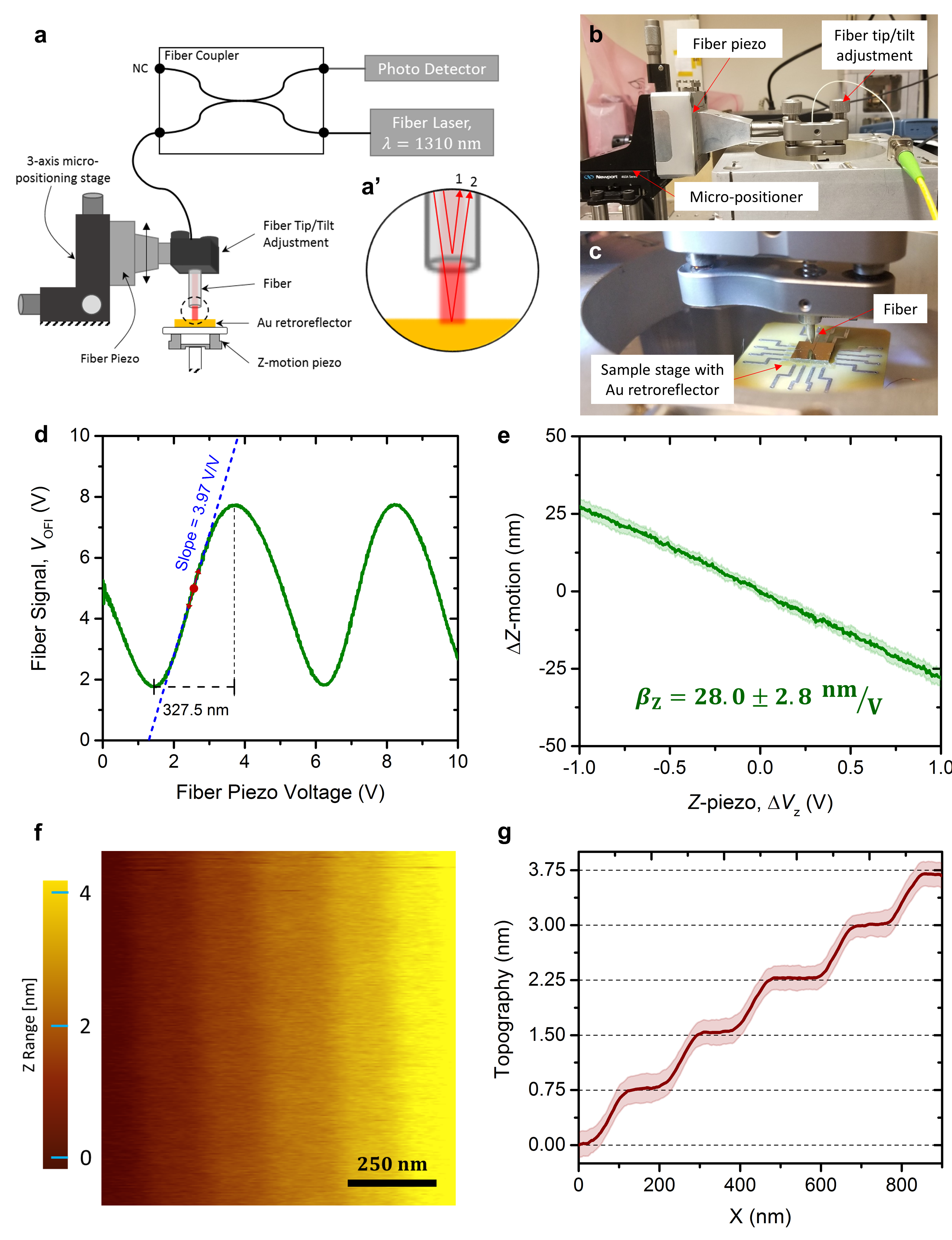}
\caption{\textbf{Calibration of the SFM \textit{z}-motion stage.} (a) Calibration setup based on an optical fiber interferometer (OFI) with $\lambda$ = 1310 nm used to calibrate the $\Delta$\textit{z}-motion of the piezo stage. (b,c) Photographs of the calibration setup. (d) Interference fringes obtained by moving the fiber position with respect to the Au retroreflector. (e) $\Delta$\textit{z}-motion of the sample stage while actuating the \textit{Z}-piezo $\Delta$\textit{V}$_{\mathrm{z}}$. Shown is the 95\% confidence interval of 20 measurements (light green) and their average (dark green). (f) SFM topographic imaging (256$\times$256 pixels) on a 6H-SiC half-monolayer test sample. (g) Statistical line trace, showing that after calibration the half-monolayer thickness corresponds to its expected value of 7.5 \AA. The solid red line is the average value of 256 lines where the shaded region represents the $\pm$1$\sigma$ uncertainty of 1.8 \AA.
\label{Fig:Z_Cal}}
\end{figure}

In order to accurately calibrate the vertical displacement of the piezo-actuated sample stage, we measure the $\Delta$\textit{z}-sensitivity of the sample stage by implementing optical fiber interferometry (OFI). Fig. \ref{Fig:Z_Cal}(a) shows the schematic of the OFI at the operating wavelength of $\lambda$ = 1310 nm, configured on top of the sample stage. For calibration, a gold (Au) retroreflector is mounted on the stage. The fiber orientation is then aligned to make normal light incidence to the retroreflector surface. After interaction with the fiber-sample junction, the reflected signals are directed towards a photodetector (Femto, OE-200-IN1). As shown in Fig. \ref{Fig:Z_Cal}(a'), a Fabry-P\'erot cavity is formed between the fiber-air interface (ray 1) and the Au retroreflector (ray 2), yielding the OFI signal highly sensitive to the $\Delta$\textit{z}-motion of the sample stage \cite{Rugar1989}. A photograph of the OFI setup is shown in Figs. \ref{Fig:Z_Cal}(b) and (c).

The first step is to calibrate the OFI sensitivity by correlating the optical fiber signal ($V_{\mathrm{OFI}}$) to vertical displacement of the fiber aperture. After the initial alignment of the optical fiber, the Fabry-P\'erot interference fringes are measured as the piezo-actuator (i.e., fiber piezo) slowly moves the fiber aperture down to the fixed Au retroreflector. As shown in Fig. \ref{Fig:Z_Cal}(d), the vertical displacement of the fiber aperture by $\lambda$/4 (i.e., 327.5 nm) should yield half a period of $V_{\mathrm{OFI}}$ oscillation due to optical interference between ray 1 and ray 2 at the fiber aperture. When the fiber aperture position slightly changes around the mid-fringe value at $V_{\mathrm{OFI}}$ = 5 V (red dot), $V_{\mathrm{OFI}}$ can be approximated to be linearly proportional to the fiber piezo voltage with a sensitivity of $\Delta V_{\mathrm{OFI}}/\Delta V_{\mathrm{FP}}$  = 3.97 V/V. Thus, the OFI sensitivity ($\beta_{\mathrm{OFI}}$) can be defined as:
\begin{equation}
    \dfrac{1}{\beta_{\mathrm{OFI}}} = \dfrac{\Delta V_{\mathrm{FP,\lambda/4}}}{\lambda/4}\dfrac{\Delta V_{\mathrm{OFI}}}{\Delta V_{\mathrm{FP}}},
\end{equation} where $\Delta V_{\mathrm{FP,\lambda/4}}$  = 2.21 V is the fiber piezo voltage required to move the fiber position by $\lambda$/4 . For our setup, $\beta_{\mathrm{OFI}}$ is determined to be 37.3 nm/V. 

Once $\beta_{\mathrm{OFI}}$ is determined, the $\Delta$\textit{z}-motion of the sample stage can be optically calibrated by $\Delta z = \beta_{\mathrm{OFI}} \Delta V_{\mathrm{OFI}}$. Fig. \ref{Fig:Z_Cal}(e) presents the $\Delta$\textit{z}-motion measured by the OFI as the \textit{z}-piezo voltage of the sample stage is varied by $\Delta V_{\mathrm{z}} = \pm$1 V about the center position at which the optical fiber aperture is initially positioned at $V_{\mathrm{OFI}} =$ 5 V. The 95\% confidence interval of 20 measurements (light green) and their average (dark green) show linear motion of the sample stage. From these measurements, the \textit{z}-piezo sensitivity about its center position is determined to be $\beta_{\mathrm{z}} = \Delta z/\Delta V_{\mathrm{z}} = $ 28.0 $\pm$ 2.8 nm/V. To verify the calibrated $\Delta$\textit{z}-sensitivity, we conducted HV-SFM imaging of a 6H-SiC half-monolayer sample whose steps are well defined as 7.5 \AA ~(K-Tek Nanotechnology). As shown in Figs. \ref{Fig:Z_Cal}(f) and (g), the calibrated $\Delta z$-sensitivity (i.e., $\beta_{\mathrm{z}} =$ 28.0 nm/V) yields the accurate half-monolayer height of  6H-SiC in the HV-SFM image. From the statistical analysis of the obtained 6H-SiC half-monolayer topographic image, the \textit{z}-position uncertainty of the sample stage is estimated to be 1.8 \AA \hspace{1mm}.

\newpage
\clearpage

\subsection*{B. Calibration of the QTF oscillation amplitude.}

\begin{figure}[h!]
\centering
\includegraphics[width=\linewidth]{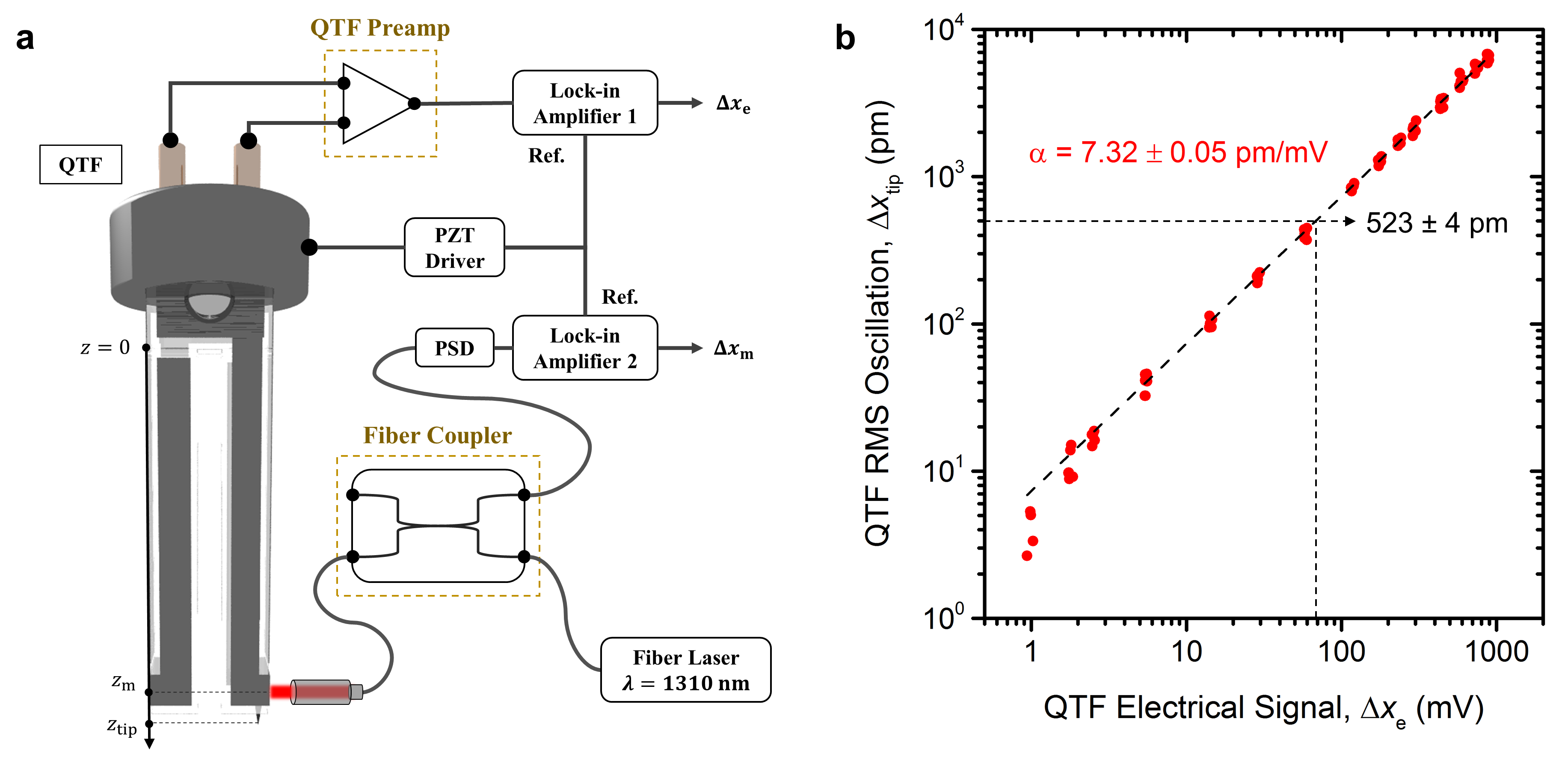}   
\caption{\textbf{QTF oscillation characterization using OFI.} (a) Implementation of the OFI system with the QTF for accurate measurement of the QTF oscillation. The QTF geometry was used to linearly relate the measured oscillation at \textit{z}$_{\mathrm{m}}$ to the tip motion at \textit{z}$_{\mathrm{tip}}$. (b) Relation between the RMS electrical QTF signal and the tip RMS oscillation in picometers. The sensitivity of the QTF signal to lateral deformation in the tines was determined to be 7.32 $\pm$ 0.05 nm/V. The QTF was nominally operated at an RMS electrical signal of $\sim$70 mV leading to an RMS tip oscillation of $\sim$500 pm.
\label{Fig:QTF_Cal}}
\end{figure}

\newpage
\clearpage

\subsection*{C. Feedback-controlled platinum nanoheaters}

\begin{figure} [h!]
\centering
\includegraphics[width=0.75\linewidth]{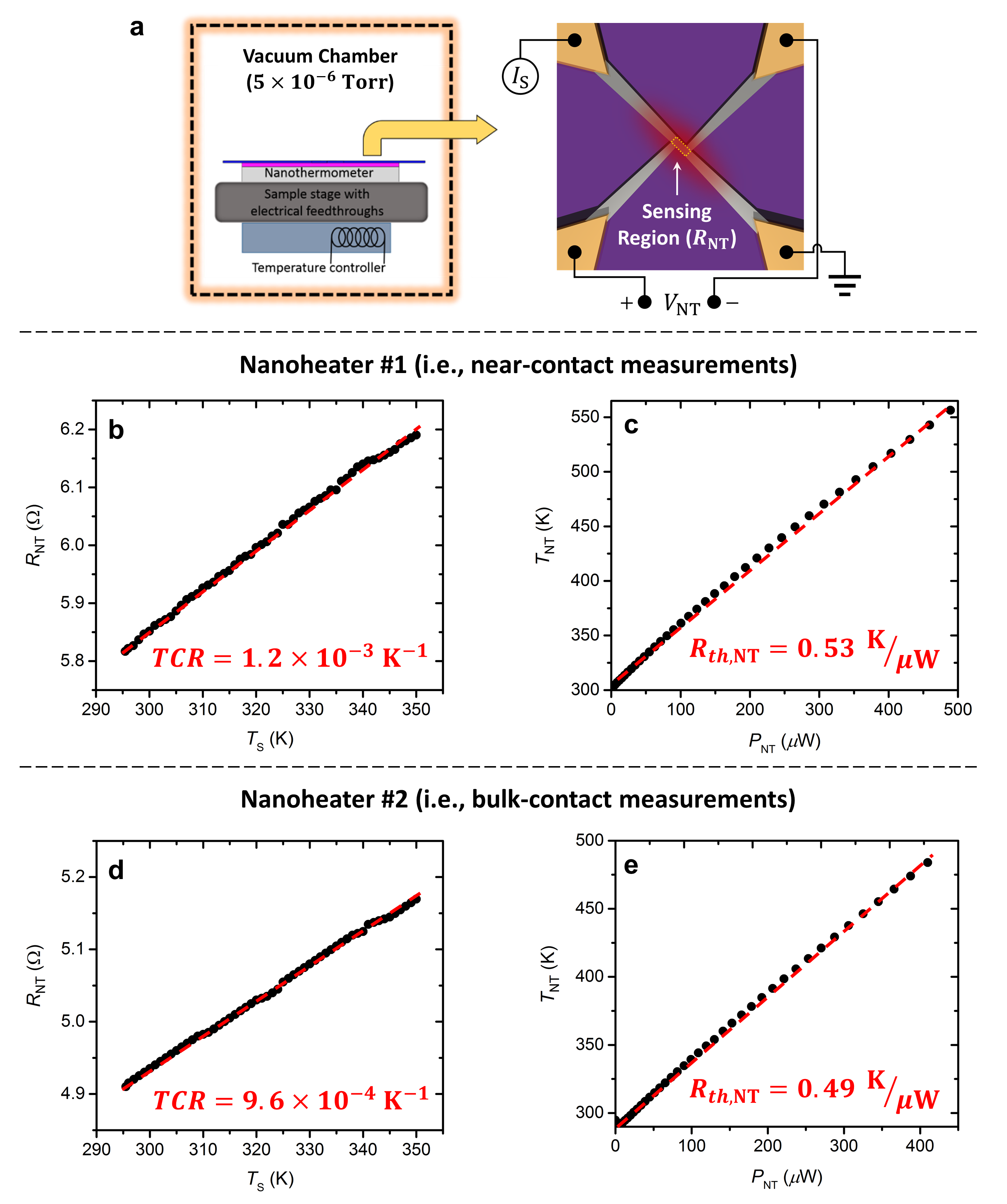}
\caption{\textbf{Calibration of the nanoheater devices.} (a) Schematic of the calibration setup for the nanoheaters. (b) Temperature coefficient of resistance (TCR) calibration for nanoheater \#1 showing a linear relationship between the nanoheater sensing region resistance (\textit{R}$_{\mathrm{NT}}$) and the substrate temperature (\textit{T}$_{\mathrm{S}}$). (c) Calibration of the sensing region thermal resistance for nanoheater \#1 (\textit{R}$_{\mathrm{th,NT}}$) by relating its temperature with its power dissipation. (d,e) Same as (b,c) but for nanoheater \#2.
\label{Fig:NH_Cal}}
\end{figure}

\newpage
\clearpage

\begin{figure}
\centering
\includegraphics[width=0.9\linewidth]{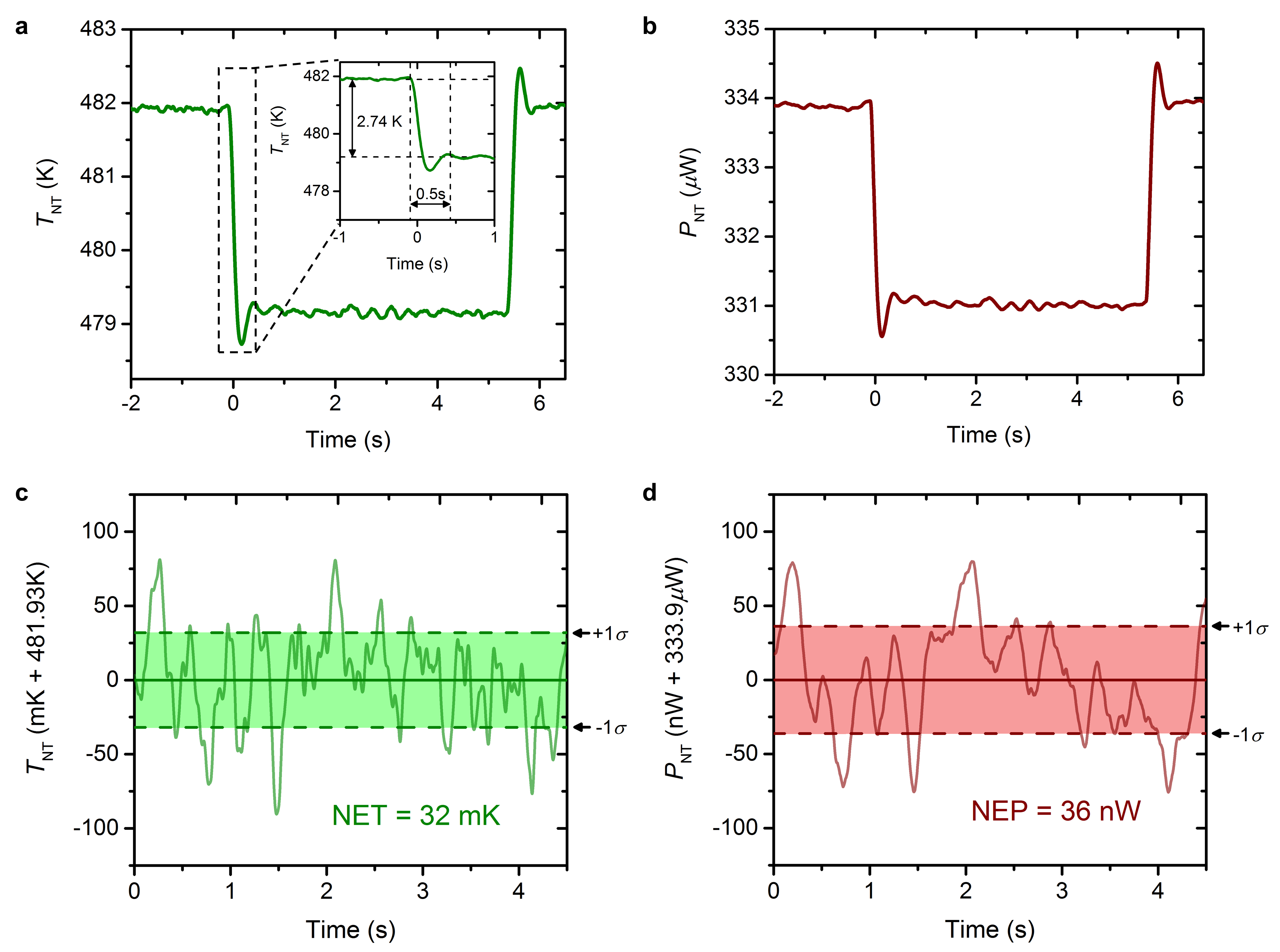}
\caption{\textbf{Characterization of the feedback-controlled nanoheater operation and its DC noise.} (a) Transient response of \textit{T}$_{\mathrm{NT}}$ during temperature feedback control (\textit{G} $=$ 20 V/$\Omega$-s) when the temperature setpoint is step-wise changed between 481.93 K and 479.19 K. (inset) shows the applied temperature change of 2.74 K for the cooling case, which requires 0.5 s to fully settle. (b) The corresponding \textit{P}$_{\mathrm{NT}}$ measurement required to stabilize the temperature at the setpoint. (c,d) Statistical analysis implemented to determine the (c) noise-equivalent-temperature (NET) and (d) noise-equivalent-power (NEP) resolutions for DC operation of the nanoheater under temperature feedback control. The $\pm$1$\sigma$ NET and NEP were determined to be 32 mK and 36 nW, respectively.
\label{Fig:NH_Char}}
\end{figure}

\newpage
\clearpage

\subsection*{D. Determination of heater lead contribution to $Q_{\mathrm{tip}}$}

\begin{figure}
\centering
\includegraphics[width=0.8\linewidth]{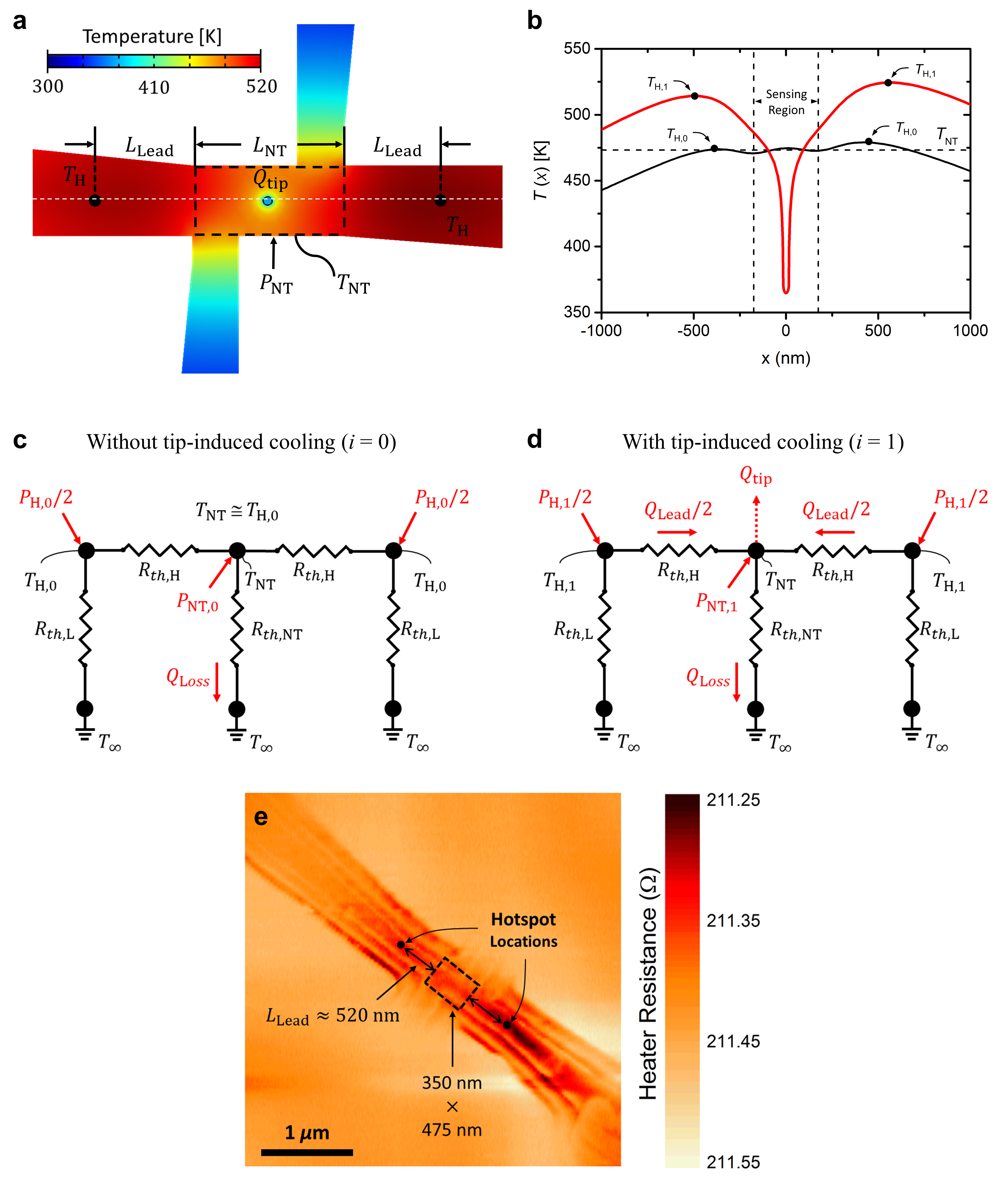}
\caption{\textbf{Analysis of heater lead contribution.} (a) Simulated temperature distribution of the nanoheater showing the characteristic temperatures and definitions used in the network analysis. (b) Surface line profile temperature extracted from the white dashed line in (a). (c,d) Thermal circuit analysis of the nanoheater without (\textit{i} $=$ 0) and with (\textit{i} $=$ 1) tip contact used to derive Eq. (\ref{Eq:Q_tip}). (e) Experimental heater resistance image acquired as a cool Si tip is raster scanned over the hot nanoheater without the use of temperature feedback control. 
\label{Fig:Sidehotspot_Cal}}
\end{figure}

Since the electrical leads outside the nanoheater sensing region are also Joule-heated during feedback-controlled nanoheater operation, its effect onto the tip-induced heat transfer measurement should be carefully characterized. Figures \ref{Fig:Sidehotspot_Cal}(a) and (b) show the numerically calculated temperature distribution (COMSOL Multiphysics) of a feedback-controlled nanoheater when a conical silicon tip with 15 nm contact radius is located at the center of the nanoheater sensing region. Although the feedback controller maintains the average volumetric temperature of the sensing region at $T_{\mathrm{NT}}$, upon the contact of the tip, tip-induced local cooling gives rise to a reversed temperature gradient (i.e., $T_{\mathrm{H}} - T_{\mathrm{NT}}$) around the sensing region (Fig. \ref{Fig:Sidehotspot_Cal}(b)). 

Figures \ref{Fig:Sidehotspot_Cal}(c) and (d) illustrate the effective thermal circuits without ($i=0$) and with ($i=1$) tip-induced cooling, respectively. When the tip is not present ($i=0$), $T_{\mathrm{H},0} \approx T_\mathrm{NT}$, and the energy balance for the sensing region can be written as
\begin{equation}\label{EB1_eq}
	P_\mathrm{NT,0} \ = \ Q_\mathrm{loss},
\end{equation}
where $Q_\mathrm{loss}$ represents the heat loss from the nanoheater sensing region to the ambient at $T_\infty = 295\ \mathrm{K}$. For $i=1$, the modified temperature profile due to tip cooling provides heat transfer avenues into the sensing region from the two external hotspots (Fig. \ref{Fig:Sidehotspot_Cal}(d)). Considering these additional heat transfer mechanisms and the symmetry of the nanoheater device, the energy balance for the sensing region can be written as
\begin{equation}\label{EB2_eq}
	P_\mathrm{NT,1} \ + \ Q_\mathrm{lead} \ - \ Q_\mathrm{tip} \ - \ Q_\mathrm{loss} \ = \ 0.
\end{equation}
When $T_{\mathrm{NT}}$ is constant under temperature feedback control, $Q_\mathrm{loss}$ is the same for both cases. Therefore, subtracting Eq. (\ref{EB1_eq}) from Eq. (\ref{EB2_eq}) yields:
\begin{equation}\label{Q_tip1}
	Q_\mathrm{tip} \ = \ \Delta P_\mathrm{NT} \ + \ Q_\mathrm{lead}.
\end{equation}
Eq. (\ref{Q_tip1}) illustrates the role of $Q_\mathrm{lead}$ to the overall value of $Q_\mathrm{tip}$. 

Under the assumption that the nanoheater is symmetric and electrical power is applied to the $T_{\mathrm{NT}}$ and $T_{\mathrm{H}}$ points, $Q_{\mathrm{lead}}$ can be approximated as:
\begin{equation}\label{Q_Lead1}
	Q_{\mathrm{lead}} \ = \ 2 \frac{(T_{\mathrm{H,1}} \ - \ T_\mathrm{NT})} {R_{th,\mathrm{H}}},
\end{equation}
where $R_{th,\mathrm{H}}$ is the conductive thermal resistance from the lead hotspot to the nanoheater sensing region. By considering both thermal circuits, the power supplied to the hotspots in the electrical leads can be written as
\begin{equation}\label{P_H}
	P_{\mathrm{H},i} \ = 2\left[\ \frac{(T_{\mathrm{H},i} \ - \ T_\mathrm{\infty})} {R_{th,\mathrm{L}}} \ + \ \frac{(T_{\mathrm{H},i} \ - \ T_\mathrm{NT})}{R_{th,\mathrm{H}}}\right],
\end{equation}
where $R_{th,\mathrm{L}}$ is the thermal resistance from the heater region to the ambient. In general, the majority of electrical power dissipation in the heater legs conducts through the substrate to ambient (i.e., $R_{th,\mathrm{L}} \ll R_{th,\mathrm{H}}$), such that Eq. (\ref{P_H}) can be reduced to $P_{\mathrm{H},i} \approx 2(T_{\mathrm{H},i} - T_\mathrm{\infty})/R_{th,\mathrm{L}}$. Therefore, it is convenient to define the ratio of heater lead power dissipation with and without tip contact:
\begin{equation}\label{P_H ratio}
	\frac{P_{\mathrm{H},1}}{P_{\mathrm{H},0}} \ \approx \ \frac{(T_{\mathrm{H},1} \ - \ T_\mathrm{\infty})} {(T_{\mathrm{H},0} \ - \ T_\mathrm{\infty})},
\end{equation}
and is valid when $R_{th,\mathrm{L}}$ does not significantly vary with temperature. For $i=0$, the lateral temperature profile is nearly uniform along the sensing region (i.e., $T_{\mathrm{H},0} \approx T_{\mathrm{NT}}$) as shown in Fig. \ref{Fig:Sidehotspot_Cal}(a). Therefore, Eq. (\ref{Q_Lead1}) can be rearranged to yield 
\begin{equation}\label{Q_Lead2}
	Q_{\mathrm{lead}} \ = \ 2 \frac{(T_{\mathrm{NT}} \ - \ T_\mathrm{\infty})} {R_{th,\mathrm{H}}}\left[\frac{P_{\mathrm{H},1}}{P_{\mathrm{H},0}} \ - \ 1\right],
\end{equation}
by combining Eq. (\ref{Q_Lead1}) and Eq. (\ref{P_H ratio}). It should be noted that $P_\mathrm{H,i}$ can be obtained by subtracting $P_\mathrm{NT}$ from the power dissipation of the entire nanoheater strip with and without tip-induced cooling. 

To quantify $Q_{\mathrm{lead}}$, $R_{th,\mathrm{H}}$ should be determined. Under the assumption that heat conduction from the hotspot to the sensing area is dominated by the Pt nanowire, $R_{th,\mathrm{H}}$ can be simplified as $R_{th,\mathrm{H}}=L_\mathrm{lead}/(\kappa A)$, 
where $\kappa$ is the thermal conductivity of the Pt nanowire, $A$ is its cross-sectional area of the nanoheater sensing region, and $L_{\mathrm{lead}}$ is the effective length of the lead from the outer edge of the inner electrode to the location at $T_{\mathrm{H}}$. Since the Pt nanowire has a sub-100 nm thickness, heat conduction is prone to size effects that reduce $\kappa$ from its bulk value. To consider these effects in the determination of $\kappa$, the Wiedeman-Franz law is used to relate the Pt nanowire's electrical properties, which are readily measured, to its thermal properties. From the Wiedemann-Franz law, the thermal conductivity is expressed as $\kappa=\sigma\mathcal{L}_{0}T_{\mathrm{NT}}$, where $\sigma$ is the electrical conductivity and $\mathcal{L}_{0}=2.44\times10^-8\ \mathrm{W \Omega/K^{2}}$ is the Lorentz number. In addition, the electrical conductivity of the Pt nanowire can be written in terms of the sensing region geometry and electrical resistance, $R_{\mathrm{NT}}$, as $\sigma \ = L_{\mathrm{NT}}/(R_{\mathrm{NT}}A)$, where $L_\mathrm{NT}$ is the length of the nanoheater sensing region. Therefore, $R_{th,\mathrm{H}}$ can be rearranged as:
\begin{equation}\label{R_H1}
	R_{th,\mathrm{H}} \ = \ \left(\frac{L_\mathrm{lead}}{\mathcal{L}_{0}T_\mathrm{NT}}\right)\left(\frac{R_{\mathrm{NT}}}{L_\mathrm{NT}}\right).
\end{equation}
Combining Eqs. (\ref{Q_Lead2}) and (\ref{R_H1}) yields the final equation for $Q_\mathrm{tip}$ as:
\begin{equation}\label{Eq:Q_tip}
	Q_\mathrm{tip} \ = \Delta P_{\mathrm{NT}}+ \frac{2\mathcal{L}_{0} T_\mathrm{NT} L_\mathrm{NT}(T_{\mathrm{NT}} \ - \ T_\mathrm{\infty})}{L_\mathrm{lead} R_{\mathrm{NT}}}\left[\frac{P_{\mathrm{H},1}}{P_{\mathrm{H},0}} \ - \ 1\right].
\end{equation}In Eq. (\ref{Eq:Q_tip}), all variables except $L_{\mathrm{lead}}$ are readily determined from the nanoheater geometry and tip-approaching experiments. $L_{\mathrm{lead}}$ is determined by imaging the electrical resistance of the entire nanoheater while an Si cantilever tip raster scans the heated nanoheater without feedback control in contact mode \cite{Hamian2016b}. Figure \ref{Fig:Sidehotspot_Cal}(e) shows the tip-scanned heater resistance image when the nanoheater is heated at $T_{\mathrm{NT}} =$ 467.25 K, demonstrating that the heater resistance becomes the lowest when the tip is positioned at the local hotspot of the lead due to the maximum tip-induced cooling. In this image, the locations of the heater lead hotspots are determined to be approximately 520 nm away from the outer edge of the inner electrodes (i.e., $L_{\mathrm{lead}} \approx$ 520 nm).

\newpage
\clearpage

\subsection*{E. Calculation of temperature rise in the tip apex}

\begin{figure} [p!]
\centering
\includegraphics[width=\linewidth]{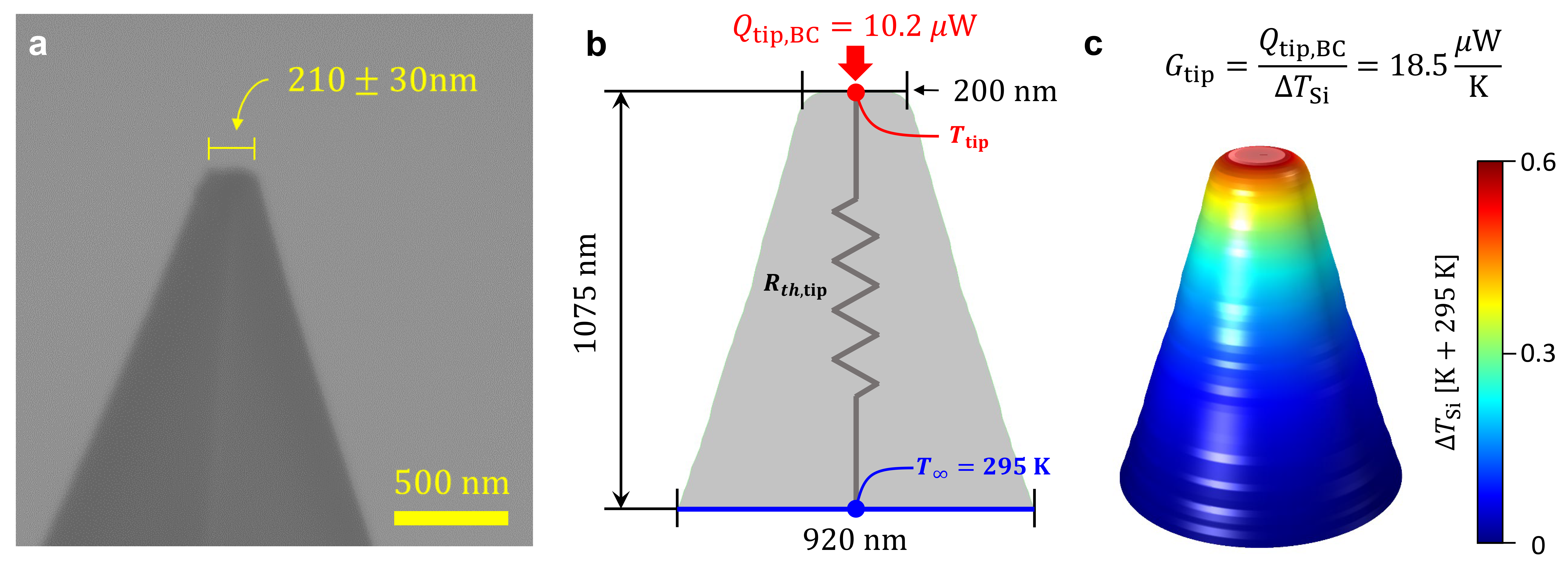}
\caption{\textbf{Calculation of the tip apex temperature.} (a) SEM image of the Si tip used for the near-contact measurements as shown in Fig. 1(c) of the main text. (b) The geometry of the Si tip in (a) is reproduced in COMSOL Multiphysics with the shown dimensions and boundary conditions. The tip's backside, located a distance of 1 $\mu$m away from the tip apex, is set to room temperature ($T_{\infty}=295\ \mathrm{K}$), while the tip apex experiences the BC heat transfer rate of $Q_{\mathrm{tip,BC}} = 10.2\ \mu\mathrm{W}$. (c) The simulated temperature distribution in the Si tip. The tip apex temperature is elevated by 0.6 K due to $Q_{\mathrm{tip,BC}}$, which is negligible compared to the temperature gradient of the experiments (i.e., 172 K).
\label{Fig:Tip_Char}}
\end{figure}

To define $G_{\mathrm{exp}}$ in both the near-contact and bulk-contact experiments, the temperature gradient across the nanogap/interface was defined as $\Delta T= T_{\mathrm{NT}}-T_{\mathrm{tip}}$. While $T_{\mathrm{NT}}$ is readily determined using the feedback-controlled nanoheaters, the direct measurement of $T_{\mathrm{tip}}$ requires a functionalized thermocouple tip \cite{Kim2012a}. Since it is challenging to integrate such devices with the QTF, we have made use of COMSOL Multiphysics simultations to predict the temperature rise in the Si tip due to the heat transfer rate experienced during bulk-contact ($Q_{\mathrm{tip,BC}}= 10.2\ \mu\mathrm{W}$ at $d=0$). The geometric profile of the tip used in the near-contact measurements is obtained using SEM (Fig. \ref{Fig:Tip_Char}(a)). This profile is traced and revolved about its central axis to give a conical representation of the tip for simulation (Figs. \ref{Fig:Tip_Char}(b) and (c)). The backside of the tip is assumed at room temperature ($T_{\infty}=295\ \mathrm{K}$), while the side walls are insulated. The tip thermal conductivity is taken from intrinsic Si which does not vary significantly with doping levels less than $3\times10^{19}\ \rm cm^{-3}$ \cite{Asheghi2002}. Figure \ref{Fig:Tip_Char}(c) shows the simulated temperature distribution of the tip, where the tip apex temperature is elevated by $\Delta T_{\mathrm{Si}} = 0.6\ \mathrm{K}$ above $T_{\infty}$. The tip thermal conductance can be calculated by $G_{\mathrm{tip}} = Q_{\mathrm{tip,BC}}/\Delta T_{\mathrm{Si}}=18.5\ \mu\mathrm{W}/\mathrm{K}$, which is more than two orders of magnitude greater than the BC thermal resistance of $G_{\mathrm{BC}}\approx 58.9\ \mathrm{nW/K}$ measured in the experiments. Due to the significantly large $G_{\mathrm{tip}}$ compared to $G_{\mathrm{BC}}$, the temperature rise in the tip apex during the experiments is ignored, such that $T_{\mathrm{tip}}\approx295\ \mathrm{K}$.

\newpage
\clearpage

\newpage
\clearpage

\subsection*{F. Experimental setup for bulk-contact thermal conductance measurement}

\begin{figure} [h!]
\centering
\includegraphics[width=\linewidth]{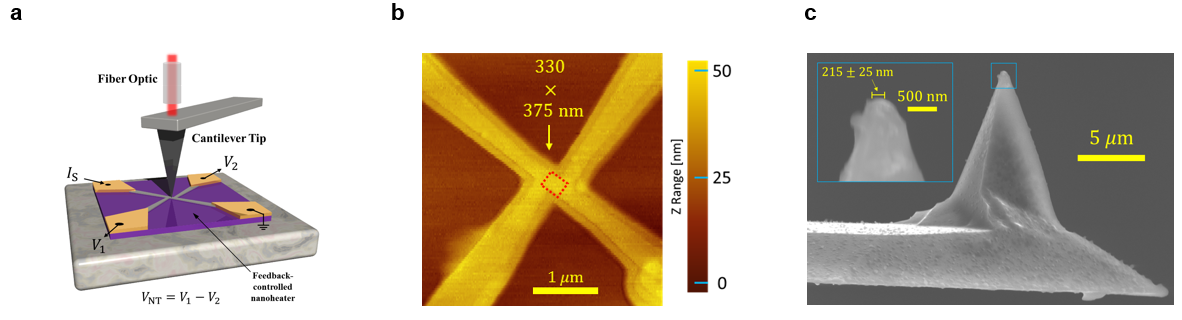}
\caption{(a) Experimental setup, where the Si tip is mounted to the free end of a microcantilever whose deflection is detected by a fiber optic positioned above it. (b) AFM topographic image of the nanoheater used for the BC experiment with a sensing area of 330 nm $\times$ 375 nm, as marked by a red rectangle. (c) SEM image of the cantilever probe tip used for the BC experiment, whose flattened contact diameter is 215$\pm$25 nm.
\label{Fig:BC_Expsetup}}
\end{figure}

\newpage
\clearpage

\newpage
\clearpage

\section*{II. Verification of the 1D AGF method}
\begin{figure}
    \centering
    \includegraphics[width=0.8\linewidth]{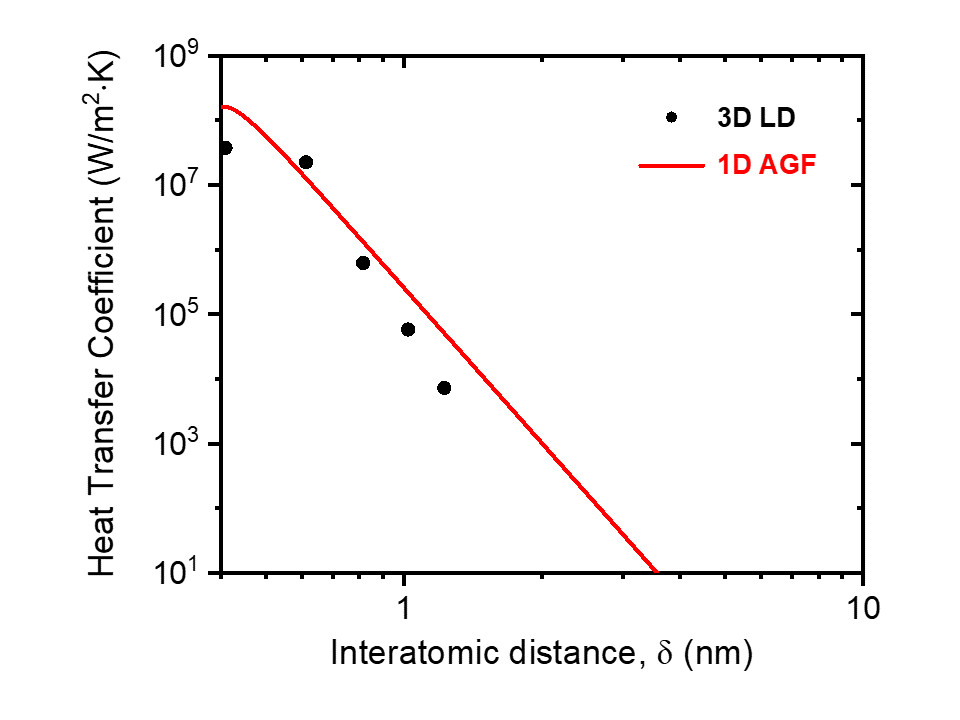}
    \caption{$\bm{\vert}$ \textbf{Comparison between the 1D AGF method used in this work and 3D LD simulations from Ref. \cite{Alkurdi2020}.} Both calculations consider two gold surfaces separated by an interatomic vacuum distance, $\delta$.}
    \label{fig:1D_3D}
\end{figure}

The acoustic phonon heat transfer coefficient calculated with the 1D AGF method is compared against previous theoretical and experimental results for verification purpose. The acoustic phonon heat transfer coefficient between gold surfaces calculated with the 1D AGF at 300 K using the Lennard-Jones potential with the parameters obtained from Ref. \cite{Yu2004} is shown in Fig. \ref{fig:1D_3D} and is compared against three-dimensional (3D) lattice dynamics (LD) results from Ref. \cite{Alkurdi2020}. The good agreement between these results suggest that a 1D approximation for acoustic phonon transport is sufficient to describe near-contact heat transfer.

Table \ref{tab:Comparison} provides Si-Si interfacial heat transfer coefficients obtained via different models, including the diffuse mismatch model (DMM) \cite{Landry2009},  molecular dynamics (MD) \cite{Aubry2008}, the 3D LD \cite{Sellan2012}, the 3D AGF \cite{Zhang2007e}, and the 1D AGF with the Stillinger-Weber potential (this work). Experimental interfacial heat transfer coefficient from Ref. \cite{Schroeder2015} is also listed, where the lower and upper limits come from the different terminating states of the Si atoms: (lower) between hydrogen/hydrogen-terminated and (upper) between hydrogen/oxygen-terminated interfaces. Table \ref{tab:Comparison} confirms a good agreement between the 1D AGF method with previous theoretical and experimental Si-Si interfacial heat transfer coefficients. 

\newpage
\clearpage

\begin{center}
    \begin{table}
    \setcounter{table}{0}                     
    \renewcommand{\thetable}{S\arabic{table}} 
        \centering
        \caption{$\bm{\vert}$ \textbf{Comparison of the Si-Si interfacial heat transfer coefficient ($h_{\mathrm{Si\text{-}Si}}$) calculated using the 1D AGF method against results reported in the literature.}}
        \vspace{12pt}
        \begin{tabular}{ccc}
            \hline
            & \hspace{0.5cm} $h_{\mathrm{Si\text{-}Si}}$ ($\times$10$^8$ W/m$^{2}$${\cdot}$K) \hspace{0.5cm} & \hspace{0.5cm} Relative Difference (\%) \hspace{0.5cm} \\ \hline
            DMM \cite{Landry2009} & 7.28 & 15.8 \\
            MD \cite{Aubry2008}& 7.69 & 11.1 \\
            3D LD ($\delta=$ 0.1 nm) \cite{Sellan2012} & 5.51 & 36.3 \\
            3D AGF \cite{Zhang2007e} & 9.09 & 5.1\\ 
            Experiment \cite{Schroeder2015} & 1.09 - 7.69 & 87.4 - 11.1 \\ \hline
            1D AGF [This work] & 8.65 & $-$ \\ \hline
        \end{tabular}
        \label{tab:Comparison}
    \end{table}
\end{center}

\newpage
\clearpage

\newpage
\clearpage

\section*{III. Effect of bias voltage on tip-sample lateral force} \label{Secion:Bias}

\begin{figure}
\centering
\includegraphics[width=1.0\linewidth]{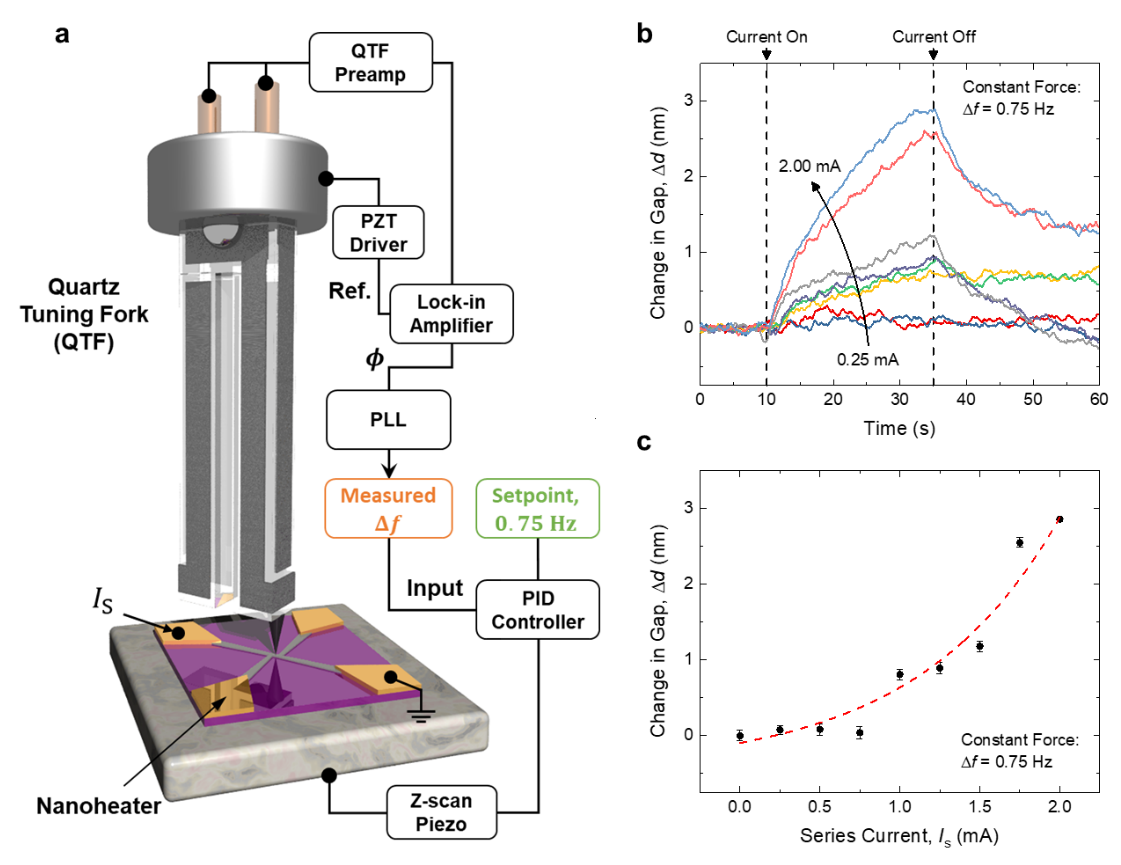}
\caption{\textbf{Measured effect of the electronic nanoheater operation on the tip-sample force.} (a) Experimental schematic based on a quartz tuning fork (QTF) and nanoheater to measure electrostatic forces induced by the series current, $I_{\rm s}$. The QTF resonant frequency shift, $\Delta f$, due to tip-sample forces is monitored electrically. $\Delta f$ is maintained constant at 0.75 Hz by closed-loop feedback control of the sample stage’s z-piezo. (b) Change in gap distance, $\Delta d$, as $I_{\rm s}$ is temporally varied from 0.25 to 2.00 mA in increments of 0.25 mA. The current is turned on at 10 s, causing $\Delta d$ to increase. At 35 s, the current is turned off, resulting in a decrease of $\Delta d$. (c) Maximum value of $\Delta d$ extracted from (b) between 33-35 s as a function of $I_{\rm s}$. Red curve in (c) is used to guide the eye.\label{Fig:AdditionalExperiments}}
\end{figure}

Figure 7(b) of the Main text and the results in Ref. \cite{Ezzahri2014a} illustrate direct proportionality between acoustic phonon heat transfer and force in the NC regime. In addition, the AGF model predicts that the Coulomb force is the main contributor to enable acoustic phonon transport in the NC regime. Therefore, we can hypothesize that if the tip-sample lateral force can be controlled via an applied electrostatic field, the heat transfer can be controlled as well.
Figure \ref{Fig:AdditionalExperiments}(a) illustrates the experimental setup for the investigation of the electrostatic force effect. The experimental conditions (e.g., tip material and shape, nanoheater material, vacuum condition) are set to mimic that of the main text experiments. For this measurement, the QTF resonance frequency shift ($\Delta f$) is maintained constant in the NC regime at 0.75 Hz ($\sim$0.5 nN-rms) by closed-loop feedback control of the sample stage’s \textit{z}-piezo. This ensures a sufficient gap space that would prevent accidental contact due to parasitic drift while still being sensitive to the tip-substrate force change. Therefore, any subtle change of an electrostatic force can be monitored by tracing the vertical displacement of the piezo-driven sample stage and the resultant change in the tip-sample gap distance (i.e., $\Delta d = d_{\rm e} - d_{\rm 0}$, where $d_{\rm e}$ and $d_{\rm 0}$ are the gap distance with and without an applied electrostatic force, respectively). On the nanoheater side, a sensing current ($I_{\rm s}$) flows through heater leads, which proportionally induces a bias voltage at the nanoheater sensing region. Thus, by changing $I_{\rm s}$ it is anticipated that the tip-sample lateral force can be dynamically controlled.

The temporal measurement of $\Delta d$ is shown in Fig. \ref{Fig:AdditionalExperiments}(b), where $I_{\rm s}$ is step-wise turned on and off at 10 s and 35 s, respectively. On-values of $I_{\rm s}$ were ranged from 0.25 to 2.00 mA in increments of 0.25 mA. Here, the current was purposely kept below the value used in the main text ($\sim$6.5 mA) to prolong nanoheater usage. Clearly, when $I_{\rm s}$ is turned on, $\Delta d$ increases, and when $I_{\rm s}$ is turned off, $\Delta d$ decreases. It should be noted that the integral gain of the \textit{z}-feedback controller was set to 3.0 nm/Hz-s to preserve signal-to-noise ratio of the measurement. While this provided a low noise threshold, it reduced the system’s responsiveness to the step-wise $I_{\rm s}$ variation, such that the experiment had to be conducted over a timescale of seconds. The maximum $\Delta d$ achieved during the measurements is quantified between 33-35 s, just before the current is turned off, and plotted as a function of $I_{\rm s}$ in Fig. \ref{Fig:AdditionalExperiments}(c). Apparently, there is a nonlinear increase of $\Delta d$ with $I_{\rm s}$ (illustrated by the red dashed line in Fig. \ref{Fig:AdditionalExperiments}(c)) to a maximum value of nearly 3 nm at 2.0 mA, which indiates that the AFM feedback controller increases the gap distance to maintain $\Delta f$ (thus the tip-nanoheater force interaction) at the set-point. It should be noted that at 2.0 mA the temperature of the nanoheater is elevated ($\sim$350 K) and results in a negligible amount of thermal expansion (0.02 nm) \cite{Hamian2016b}. The obtained result suggests that increasing $I_{\rm s}$ should result in stronger electrostatic forces between the Si tip and Pt nanoheater, which we believe leads to the increase of the acoustic phonon transport.


\newpage
\clearpage

\newpage
\clearpage

\newpage
\clearpage




\newpage
\clearpage
